 \documentclass[twocolumn]{aastex631}

\shorttitle{Multiwavelength Analysis of NGC 449}
\shortauthors{Guo et al.}
\graphicspath{{./}{figures/}}
\usepackage{amsmath}	
	
\usepackage{multirow}
\usepackage{threeparttable}

\usepackage{CJKutf8}

\begin{document}

\title{Multiwavelength Analysis of a Nearby Heavily Obscured AGN in NGC 449}

\correspondingauthor{Qiusheng Gu}
\email{qsgu@nju.edu.cn}

\author[0000-0002-2338-7709]{Xiaotong Guo \begin{CJK*}{UTF8}{gkai}(郭晓通)\end{CJK*}}
\affiliation{Institute of Astronomy and Astrophysics, Anqing Normal University, Anqing, Anhui 246133, China}

\author[0000-0002-3890-3729]{Qiusheng Gu \begin{CJK*}{UTF8}{gkai}(顾秋生)\end{CJK*}}
\affiliation{School of Astronomy and Space Science, Nanjing University, Nanjing, Jiangsu 210093, China}
\affiliation{Key Laboratory of Modern Astronomy and Astrophysics (Nanjing University), Ministry of Education, Nanjing 210093, China}


\author[0000-0003-1697-6801]{Jun Xu \begin{CJK*}{UTF8}{gkai}(徐骏)\end{CJK*}}
\affiliation{Institute of Astronomy and Astrophysics, Anqing Normal University, Anqing, Anhui 246133, China}

\author[0000-0001-9694-2171]{Guanwen Fang \begin{CJK*}{UTF8}{gkai}(方官文)\end{CJK*}}
\affiliation{Institute of Astronomy and Astrophysics, Anqing Normal University, Anqing, Anhui 246133, China}

\author[0000-0002-8348-2783]{Xue Ge \begin{CJK*}{UTF8}{gkai}(葛雪)\end{CJK*}}
\affiliation{School of Physics and Electronic Engineering, Jiangsu Second Normal University, Nanjing, Jiangsu 211200, China}

\author[0000-0001-5895-0189]{Yongyun Chen \begin{CJK*}{UTF8}{gkai}(陈永云)\end{CJK*}}
\affiliation{College of Physics and Electronic Engineering, Qujing Normal University, Qujing 655011, China}

\author[0000-0002-2937-6699]{Xiaoling Yu \begin{CJK*}{UTF8}{gkai}(俞效龄)\end{CJK*}}
\affiliation{College of Physics and Electronic Engineering, Qujing Normal University, Qujing 655011, China}

\author[0000-0003-1028-8733]{Nan Ding \begin{CJK*}{UTF8}{gkai}(丁楠)\end{CJK*}}
\affiliation{School of Physical Science and Technology, Kunming University, Kunming 650214, China}



\begin{abstract}

We presented the multiwavelength analysis of a heavily obscured active galactic nucleus (AGN) in NGC 449. We first constructed a broadband X-ray spectrum using the latest NuSTAR and XMM-Newton data. 
Its column density ($\simeq 10^{24} \rm{cm}^{-2}$) and photon index ($\Gamma\simeq 2.4$) were reliably obtained by analyzing the broadband X-ray spectrum. 
However, the scattering fraction and the intrinsic X-ray luminosity could not be well constrained. 
Combined with the information obtained from the mid-infrared (mid-IR) spectrum and spectral energy distribution (SED) fitting, we derived its intrinsic X-ray luminosity ($\simeq 8.54\times 10^{42} \ \rm{erg\ s}^{-1}$) and scattering fraction ($f_{\rm{scat}}\simeq 0.26\%$).
In addition, we also derived the following results:
(1). The mass accretion rate of central AGN is about $2.54 \times 10^{-2} \rm{M}_\odot\ \rm{yr}^{-1}$, and the Eddington ratio is $8.39\times 10^{-2}$; (2). The torus of this AGN has a high gas-to-dust ratio ($N_{\rm H}/A_{\rm V}=8.40\times 10^{22}\ \rm{cm}^{-2}\ \rm{mag}^{-1}$); (3). The host galaxy and the central AGN are both in the early stage of co-evolution.

\end{abstract}



\section{Introduction} \label{sec:intro}
It is well known that active galactic nuclei (AGNs) are powered by the accretion of surrounding matter by supermassive black holes (SMBHs). The emission of AGNs covers almost the whole electromagnetic band. The AGNs' radiations at different wavelengths arise from their diverse structures. For example, the significant emission of the accretion disk is in the ultraviolet (UV) to the optical band \citep[e.g.,][]{1980ApJ...235..361R}, while the emission of the torus is in the mid-infrared (mid-IR) band \citep[e.g.,][]{1993ARA&A..31..473A}. 
The diverse information can be derived from the emission of AGNs at different wavelengths.
To fully understand an AGN, it is often necessary to study with the multi-bands.

The classification of AGNs usually utilizes the observation features of different wavelengths. 
For instance, based on the characteristics of the optical emission lines, AGNs can be classified into type 1 and type 2 \citep[e.g.,][]{1974ApJ...192..581K}. The AGN unified model proposes that different AGN types are caused by different viewing angles with an obscuring torus \citep[e.g.,][]{1993ARA&A..31..473A,1995PASP..107..803U,2015ARA&A..53..365N}.  Since the X-ray emission has a strong pierce, the X-ray emission of AGNs is usually subject to little absorption. Based on the column density ($N_{\rm H}$) of the X-ray absorption, AGNs can be classified into four categories: unobscured ($N_{\rm H}<10^{22}~{\rm cm}^{-2}$), obscured ($10^{22}~{\rm cm}^{-2}<N_{\rm H}<10^{23}~{\rm cm}^{-2}$), heavily obscured ($10^{23}~{\rm cm}^{-2}<N_{\rm H}<1.5\times 10^{24}~{\rm cm}^{-2}$), and Compton-thick \citep[$N_{\rm H}\geqslant 1.5\times 10^{24}~{\rm cm}^{-2}$, e.g.,][]{2004ASSL..308..245C}.
Many studies have also regarded the source with a column density of $\sim 10^{24}~{\rm cm}^{-2}$ as Compton-thick AGN \citep[CT-AGN, e.g.,][]{1999ApJ...522..157R,2019ApJ...877..102K}.

Some studies have pointed out that most AGNs were heavily obscured \citep[e.g.,][]{2018ARA&A..56..625H}. The heavily obscured AGNs are believed to be in an early phase of the evolutionary scenario of AGNs, which is a rapid growth state of central SMBH \cite[e.g.,][]{2011MNRAS.411.1231G}. The host galaxies of heavily obscured AGNs are also in an early phase of the evolutionary scenario of galaxies. For instance, \cite{2015ApJ...814..104K} pointed out that heavily obscured AGNs were twice as likely to be hosted by late-type galaxies relative to unobscured AGNs.
As a class of the most luminous and persistent X-ray sources in the universe, AGNs contribute the vast majority of Cosmic X-ray Background \cite[CXB, e.g.,][]{2014IAUS..304..125U} in the 1 to $\sim$200 -- 300 keV.
The heavily obscured AGNs and CT-AGNs are significant contributors to the hump of the CXB emission at 30 keV \citep[e.g.,][]{1980ApJ...235....4M, 2008ApJ...689..666A}.

The single-band analysis for AGNs can only derive a part of their properties.
For example, we can derive the column density in the line-of-sight (LOS) direction by fitting X-ray spectra. Using the spectral energy distributions (SEDs) from the UV to the far-IR, we can derive the star formation rates (SFRs) of the host galaxies, stellar masses of the host galaxies, and AGN luminosities.
To fully understand heavily obscured AGNs, their host galaxies, and their co-evolution, we first derive their properties through multi-band analysis.

The heavily obscured AGNs usually present a complex spectral shape in their X-ray waveband.
Theoretically, the soft X-ray emission (E $\lesssim$ 2 keV) of heavily obscured AGNs will be heavily (entirely) absorbed. Even in many CT AGNs, the soft X-ray is not entirely absorbed, because the emission from the AGNs' scattered and the host galaxy can be observed \citep[e.g.,][]{1997ApJ...488..164T,2012ApJ...758...82L}.
As the energy increases, the absorption of X-ray emission (E$>$10 keV) is gradually reduced or cut-off. The hard X-ray of heavily obscured AGN is dominated by the reflected component, the so-called Compton reflection continuum characterized by a broad bump peaking around 20--30 keV \citep[e.g.][]{2004ASSL..308..245C}. One of the most prominent features in the X-ray spectrum of heavily obscured AGN is the neutral iron K$\alpha$ emission line at 6.4 keV. However, the X-ray spectra of some heavily obscured AGNs show a weak (or absent) iron K$\alpha$ emission line \citep{2018MNRAS.477.3775B}, especially the heavily obscured AGNs beyond the local universe.
To accurately determine the column density of the gas and recover the intrinsic X-ray flux of AGNs, we require proper modeling of the physical processes in the AGNs that self-consistently account for the transmitted emission, Compton-scattered reflected component and the fluorescent line flux. Such modeling may also constrain the geometry of the torus and view angle. Over the past several years, a series of X-ray spectral models \citep[e.g.,][]{2009MNRAS.397.1549M,2018ApJ...854...42B} are created with assumed different geometries (spherical, toroidal, or clumpy).  
Many recent studies use these models to constrain the parameters of AGNs  \citep{2017MNRAS.467.4606G,2019ApJ...877..102K,2019ApJ...887..173L,2020ApJ...888....8T}.

From the UV to the optical band, the continua of the heavily obscured AGNs are masked by the emission of the host galaxies, so we can only observe their high-ionized emission lines and obtain little information about them.  
It is difficult to constrain the gas column density along LOS for the AGNs with low-quality X-ray spectra. However, we can estimate the gas column density using the luminosity ratio of X-ray to high-ionized emission lines  \citep[e.g.,][]{1998A&A...338..781M, 2006A&A...446..459C, 2010A&A...519A..92G, 2015A&A...578A.120L}. 
The mid-IR emission of the AGNs is usually less suppressed due to the low optical depth. The contribution of AGNs can be easily decomposed in the mid-IR spectra. In addition, for heavily obscured AGNs, we can also observe a strong silicon absorption line at 9.7~\micron. The dust in the torus can be derived using silicon absorption strength \citep[e.g.,][]{2020RAA....20..147X}. 
By the SED fitting, we can accurately obtain the contribution of the AGNs, the AGN luminosities, and some parameters about their host galaxies. 
So the multi-band data can provide us with more comprehensive information on heavily obscured AGNs.

NGC 449 (also known as Mrk 1) is a nearby (z=0.01595,  D$\sim$67 Mpc) Seyfert 2 galaxy, and its center hosts a heavily obscured AGN. It was observed by the XMM-Newton X-ray telescope \citep{2001A&A...365L...1J} in 2004. \cite{2005A&A...444..119G} fitted the X-ray spectrum and constrained the column density ($N_{\rm H} > 1.1\times 10^{24}\ \rm{cm}^{-2}$) of this AGN. Moreover, some studies also derived its column density by using the ratio of X-ray to high-ionized emission lines ([O III]5007, [Ne V]3426) luminosity \citep[][]{2005ApJ...634..161H,2010A&A...519A..92G}. These results indicated that the X-ray is severely absorbed and that even the AGN may be a CT-AGN. However, these results did not constrain its column density well. To better constrain its column density and other properties, the spectrum above 10 keV is necessary. Fortunately, this source was observed by the Nuclear Spectroscopic Telescope Array \citep[NuSTAR,][]{2013ApJ...770..103H} in 2017. The NuSTAR data has still not been analyzed (or presented in previous work).
Moreover, there are abundant data about NGC 449 in the archives, which can also help us to obtain more properties. So the motivation of our work is to constrain the properties of this heavily obscured AGN by multiwavelength analysis.

The structure of the paper is as follows. Section~\ref{sec:Multi-data} describes the multiwavelength data and processing for NGC 449. In Section~\ref{sec:analysis}, we analyzed the multiwavelength data of this source in detail. Then, we discuss the implications of our results in Section~\ref{sec:result}. Finally, we present a summary of this work in Section~\ref{sec:summary}. We adopt a concordance flat $\Lambda$-cosmology with ${\rm H_0 = 67.4 \ km\ s^{-1}\ Mpc^{-1}}$, $\Omega_{\rm m} = 0.315$, and $\Omega_\Lambda = 0.685$ \citep{2020A&A...641A...6P}.

\section{Multiwavelength data}\label{sec:Multi-data}
\subsection{X-Ray Observations and Data Reduction}\label{sec:X-ray-data}
A log of the XMM-Newton and NuSTAR observations analyzed herein is presented in Table~\ref{Tab:obs}, and the individual data sets are described in this section.

\begin{deluxetable*}{lccccr}
	\label{Tab:obs}
	\tablecaption{Observation log.}
	\tablewidth{0pt}
	\tablehead{
		\colhead{ObsID} & \colhead{Observation date} & \colhead{Mission} & \colhead{Instrument(s)} &
		\colhead{Net Exposure} & \colhead{Net Counts}\\
		&&&&\colhead{[ks]}&
	}
	\decimalcolnumbers
	\startdata
	& && MOS1 &     11.4 & 249.90  \\
	0200430301&2004-01-09     &Newton  & MOS2 & 11.5& 222.31\\
	& && PN &  8.6& 735.63  \\
	\hline\noalign{\smallskip}
	\multirow{2}{*}{60360002002}&\multirow{2}{*}{2017-12-23}     &\multirow{2}{*}{NuSTAR}  & FPMA     &    32.7    &  130.66          \\
	&&& FPMB     &       32.6 &  126.67  \\
	\enddata
\end{deluxetable*}

\subsubsection{XMM-Newton Observations}
NGC 449 was observed by the XMM-Newton X-ray telescope \citep{2001A&A...365L...1J} for about 11.9 ks of exposure on January 9, 2004 (PI: Matteo Guainazzi, ObsID: 0200430301).
The data is reduced in a standard manner using the XMM-Newton Science Analysis System \citep[SAS,][]{2004ASPC..314..759G} v17.0.0 and Current Calibration Files (CCF) of June 22, 2018.
The spectra are extracted from a circular region (30 arcsec radius for PN and 25 arcsec radius for MOS) around the source, and the background spectra are taken from a nearby source-free circular region (80 arcsec radius for PN and 100 arcsec radius for MOS). The spectra are binned to have minimum counts of 10 per energy bin.

\subsubsection{NuSTAR Observations}
NGC 449 was also observed by the NuSTAR \citep{2013ApJ...770..103H} for about 32.7 ks of exposure on December 8, 2017 (ObsID: 60360002002).
The data is processed using the NuSTAR data analysis software \textit{nustardas v1.9.5} available in \textit{heasoft v6.27.2} and CALDB released on August 13, 2020. The \textit{nupipeline} script is used to produce calibrated and clean event files.
We extract source spectra using the \textit{nuproducts} task.
The spectra are extracted from a 45 arcsec radius aperture centered on the source, while the background is extracted from a circular region with a 100 arcsec radius from a source-free region on the detector.
To improve the spectral quality of NuSTAR, we try to combine the two spectra of FPMA and FPMB.
The combined spectrum is also binned to have minimum counts of 10 per energy bin.
\subsection{Mid-IR spectral Observations and Data Reduction}
We search the archives for its mid-IR data. There is good mid-IR spectral data, covering wavelengths from 5 to 37~\micron, observed by Spitzer's IrsStare on February 5, 2009 (PI: Levenson, Nancy, ObsID: 25408000).
The mid-IR spectrum was processed using the Spitzer data analysis software \textit{irs\_merge v2.1} with the \textit{pipeline} script (version: S18.18.0) on December 18, 2011. For more detailed processing of the mid-IR spectrum of this source, please refer to Section 2 of \cite{2011ApJS..196....8L}.

\subsection{Multiwavelength photometric data}
To construct the SED for NGC 449, we search for available multi-band photometric data from different surveys or the archives of telescopes, i.e., the Sloan Digital Sky Survey (SDSS) Data Release 7 \citep{2009ApJS..182..543A}, the Two Micron All-Sky Survey \citep[2MASS, ][]{2006AJ....131.1163S}, the Wide-field Infrared Survey Explorer \citep[WISE, ][]{2010AJ....140.1868W}, the AKARI and the Infrared Astronomical Satellite (IRAS). 
In total, we collect the photometric data for 17 filters. Among them, only the filter SDSS-u is in the UV band. There are three filters of the optical band, including SDSS-g, SDSS-r, and SDSS-i. The other filters are in the IR band, including four near-IR filters (SDSS-z, 2MASS-J, 2MASS-H, 2MASS-K), seven mid-IR filters (W1, W2, W3, W4, IRAS-PSC 25, AKARI-PSC 09, AKARI-PSC 18) and two far-IR filters (IRAS-PSC 60, IRAS-PSC 100).
The details for photometric data and filters are listed in Table~\ref{Tab:photometry}.

\begin{deluxetable}{lccc}
	\label{Tab:photometry}
	\tablecaption{Photometry of NGC 449.}
	\tablewidth{0pt}
	\tablehead{
		\colhead{Band} & \colhead{Wavelength} & \colhead{Flux} & \colhead{References} \\
		&\colhead{[\micron]}&\colhead{[mJy]}&
	}
	\decimalcolnumbers
	\startdata
	SDSS-u&0.3547&$1.5473 \pm 0.0109$&\multirow{5}{*}{(a)}\\
	SDSS-g&0.4767&$5.1128 \pm 0.0104$&\\
	SDSS-r&0.6226&$8.5429 \pm 0.0161$&\\
	SDSS-i&0.7615&$10.024 \pm 0.0196$&\\
	SDSS-z&0.9123&$12.909 \pm 0.0368$&\\
	\hline
	2MASS-J&1.234&$19.8 \pm 0.704$&\multirow{3}{*}{(b)}\\
	2MASS-H&1.661&$23.4 \pm 1.08$&\\
	2MASS-K&2.157&$22.6 \pm 1.11$&\\
	\hline
	WISE-W1&3.4&$13.035 \pm 0.2084$&\multirow{4}{*}{(c)}\\
	WISE-W2&4.6&$20.144 \pm 0.3220$&\\
	WISE-W3&12&$116.39 \pm 1.8618$&\\
	WISE-W4&22&$551.74 \pm 1.9548$&\\
	\hline
	IRAS-PSC 25 &25&$801.20 \pm 128.19$&\multirow{3}{*}{(d)}\\
	IRAS-PSC 60 &60&$2302.0 \pm 299.26$&\\
	IRAS-PSC 100 &100&$2851.0 \pm 342.12$&\\
	\hline
	AKARI-PSC 09&9&$101.14 \pm 1.4838$&\multirow{2}{*}{(e)}\\
	AKARI-PSC 18&18&$531.34 \pm 25.134$&\\
	\enddata
	\tablecomments{(a)  \cite{2009ApJS..182..543A}, (b) \cite{2006AJ....131.1163S}, (c) \cite{2010AJ....140.1868W}, (d) \cite{1990IRASF.C......0M}, (e) \cite{2015ApJ...814...11T}.}
\end{deluxetable}

\section{Analysis of Multiwavelength data }\label{sec:analysis}
\subsection{X-Ray Spectral Analysis}\label{sec:X-ray-analysis}
In this section, we analyze the X-ray spectra of NGC 449 in Xspec v.12.11.0 \citep{1996ASPC..101...17A}. 
First, we use phenomenological modeling to determine whether there is variation in the spectra for the same energy range of XMM-Newton and NuSTAR.
We then analyzed which components contributed to the X-ray spectra of this source.
Based on the analysis of the X-ray components, we finally use the self-consistent and physical models to fit the X-ray spectra.
All errors represent the 68.0\% confidence intervals unless otherwise stated.

\subsubsection{Testing for Variability Among Observations}\label{sec:test}
To test whether there is significant variation in the spectra of two observations within the same energy range (3--10 keV), we need to compare their spectral shapes and fluxes.
We can obtain them by fitting the X-ray spectra of this source.
However, the data for this source is poor.
To obtain more reliable properties, we use a phenomenological model to fit the spectra of XMM-Newton\footnote{The spectra of MOS1, MOS2 and PN are simultaneously fitted.} and NuSTAR within the same energy range. The phenomenological model is written (in the Xspec terminology) as follows:
$$model=phabs[1]*cflux[2]*powerlaw[3],$$
where $phabs[1]$ accounts for the Galactic absorption which is fixed at a column density of $5.16\times10^{20}\ \rm{cm}^{-2}$.

\begin{deluxetable}{lccc}
	\label{Tab:test}
	\tablecaption{Variability among observations within the same energy range.}
	\tablewidth{0pt}
	\tablehead{
		\colhead{Mission} & \colhead{log($f_{3-10\ \rm{keV}}$)} & \colhead{$\Gamma_{3-10\ \rm{keV}}$} & \colhead{$\chi^2$/dof} \\
		&\colhead{[$\rm{erg\cdot s^{-1}\cdot cm^{-2}}$]}& &
	}
	\decimalcolnumbers
	\startdata
	Newton  & $-12.85\pm 0.13$ & $-0.86\pm 0.86$&5.02/5 \\
	NuSTAR  &   $-13.06\pm 1.58$ & $0.70\pm 0.59$& 31/17\\
	\enddata
\end{deluxetable}

Table~\ref{Tab:test} lists some properties of these two spectra within the same energy range. Their fluxes are the same within the error range. The spectral shape of XMM-Newton's spectrum within the same energy range indicates that the X-ray emission of NGC 449 is seriously absorbed. Similarly, the NuSTAR observation of this source also presents a flat spectrum. The data in the same energy range are so poor that their photon indices cannot be well constrained. In short, these two observations have similar fluxes within the same energy range. 
Therefore, there is no significant variation in the X-ray emission for NGC 449 among these two observations.

For subsequent X-ray spectrum analysis, we constitute a broadband X-ray spectrum using the X-ray spectra obtained with XMM-Newton and NuSTAR, which covers the 0.35--30 keV band (PN for XMM-Newton : 0.35--10.0 keV, NuSTAR : 3.0--30.0 keV).
\subsubsection{Analysis of the basic components}\label{sec:basic}
In this section, our goal is to analyze which components contribute to the X-ray spectra of this source, providing support for subsequent model selection.

\begin{figure*}
	\includegraphics[width=0.49\linewidth]{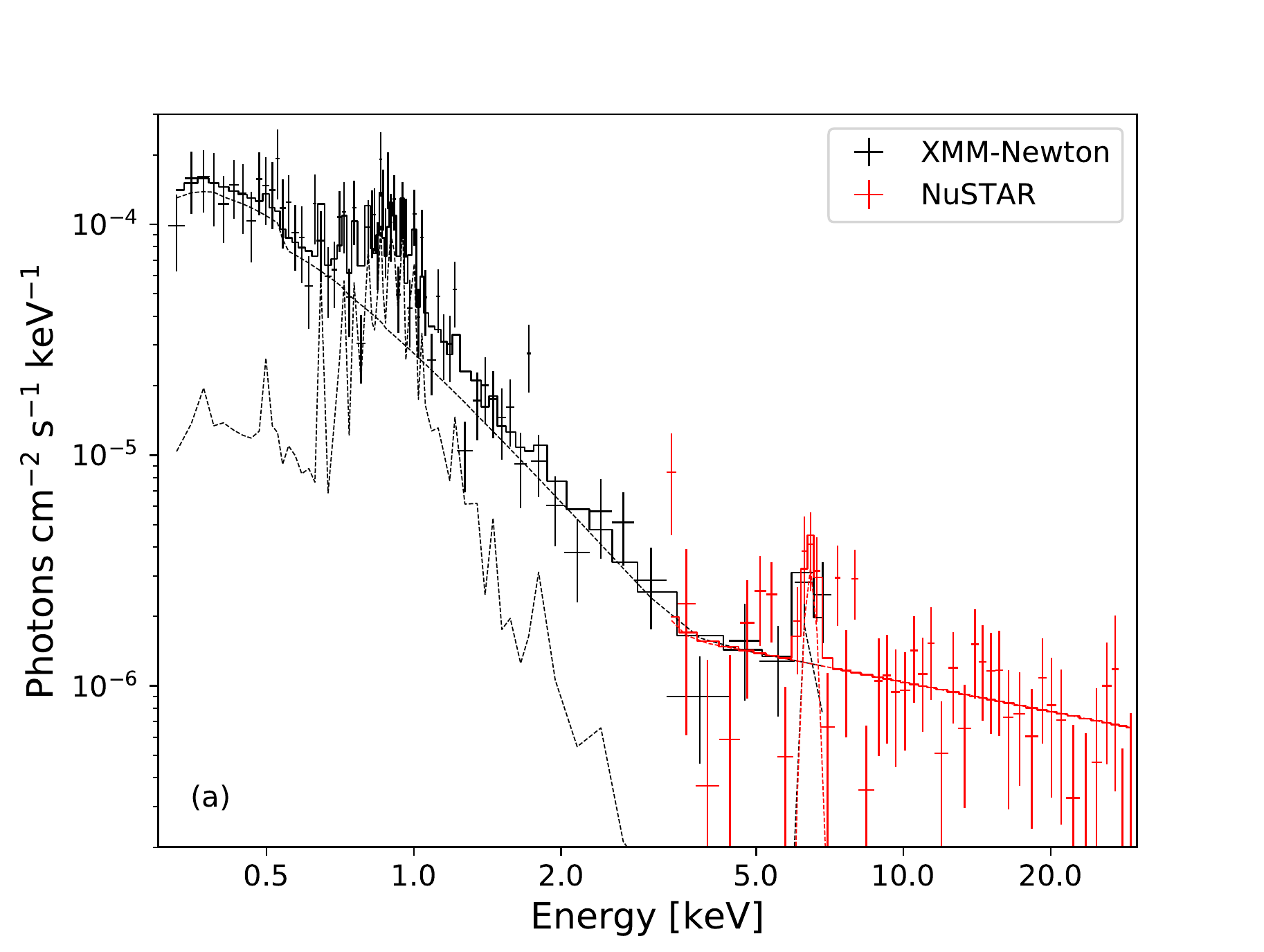}
	\includegraphics[width=0.49\linewidth]{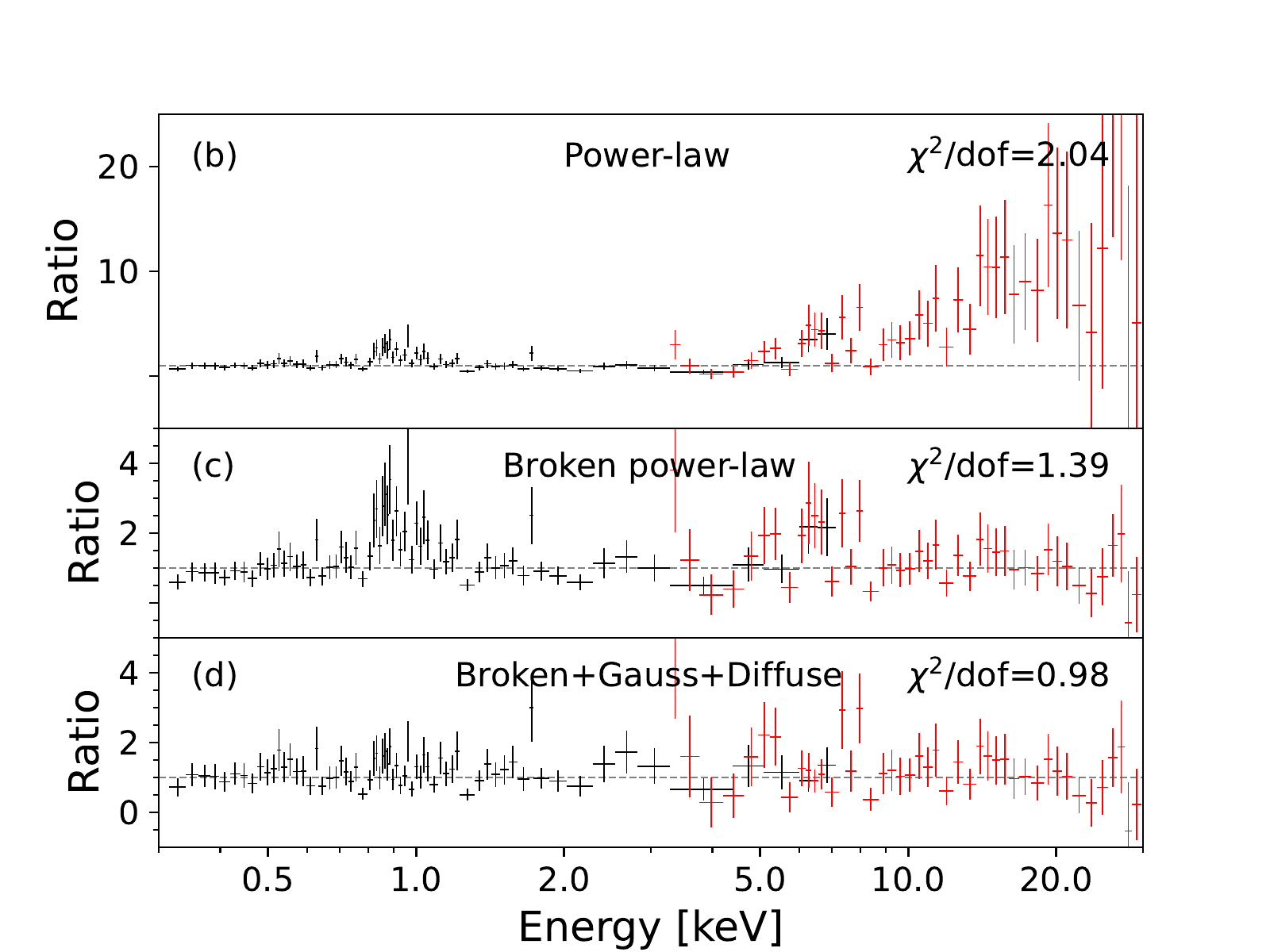}
	\caption{XMM-Newton PN (black) and NuSTAR (red) spectra of NGC 449. The black line in panel (a) represents the best-fit model ($broken+gauss+diffuse$). Panel (b) shows the ratio (data/model) by fitting the X-ray spectra using a simple power-law model. Panels (c) and (d) show the residuals obtained by fitting broken power-law and $broken+gauss+diffuse$ models, respectively.}
	\label{fig:basic}
\end{figure*}

First, We use a \textit{powerlaw} model ($phabs*powerlaw$) with a photon index of 2.17 to fit the broadband spectrum. Of course, the goodness of fit is poor (see the panel b of Figure~\ref{fig:basic}). There is significant excess in the spectrum above 10 keV, which is a flat spectrum. So the broadband spectrum is again fitted by a \textit{broken powerlaw} model ($phabs*bknpower$). Figure~\ref{fig:basic}(c) shows the residuals of the \textit{broken powerlaw} model. The best-fitting break energy is 3.93~keV. The low energy band is a steep spectrum ($\Gamma = 2.36$), and the high energy band is a flat spectrum ($\Gamma = 0.51$). 
Although its goodness of fit is significantly improved, there is still excess near 6.5~keV and 1~keV. Therefore, we also need to add two more models. Among them, a Gaussian model is used to fit the iron emission line near 6.5~keV, and the equivalent width (EW) of the iron emission line is 1.49~keV which is consistent with previous work \citep[EW$<$2~keV,][]{2005A&A...444..119G}. The other model (diffuse thermal emission) is used to match the excess near 1.0~keV. Finally, we obtain the best-fitting of the spectrum, as shown in Figure~\ref{fig:basic}(a). The fitted parameters are listed in Table~\ref{Tab:basic}.

\begin{deluxetable*}{llccc}
	\label{Tab:basic}
	\tablecaption{Analysis of the basic components.}
	\tablewidth{0pt}
	\tablehead{
		\colhead{} & \colhead{Parameter} & \colhead{Power-law} & \colhead{Broken power-law} & \colhead{Broken+Gauss+Diffuse} 
	}
	\decimalcolnumbers
	\startdata
	$powerlaw$  & $\Gamma$ & $2.17\pm 0.06$&\dots& \dots\\
	\hline
	& $\Gamma_1$ &\dots& $2.36\pm 0.08$&$2.38\pm 0.12$ \\
	$bknpower$  & $E_{Break}$ (keV)&\dots& $3.93\pm 0.50$ & $3.71\pm 0.60$ \\
	& $\Gamma_2$ &\dots& $0.51\pm 0.18$&$0.36\pm 0.19$ \\
	\hline
	\multirow{3}{*}{$zgauss$}&lineE(keV)& \dots&\dots&$6.56\pm 0.09$\\
	&$\sigma$(keV)& \dots& \dots&$0.16\pm 0.15$\\
	&EW(keV)& \dots& \dots& 1.49\\
	\hline
	$mekal$& kT(keV)& \dots &\dots &$0.82\pm 0.06$ \\
	\hline
	&$\chi^2$/dof&224.61/110 & 150.52/108 & 100.52/103 \\
	\enddata
\end{deluxetable*}

We find a flat spectrum above 3.71~keV by fitting the broadband spectrum, indicating that the emission of this source is seriously absorbed. Therefore, the reprocessed emission is dominant in hard X-rays. Theoretically, there should be a flat spectrum at energies below 3.71~keV. In fact, there is a steep spectrum below 3.71~keV with a photon index of 2.38. The power-law emission below 3.71~keV is most likely contributed by the scattered component of primary X-ray emission. Besides, the diffuse thermal emission and the iron emission line are required. 

\subsubsection{Pexmon model}
According to the analysis in Section~\ref{sec:basic}, we fit the spectrum using the neutral reflection model \textit{Pexmon} \citep{2007MNRAS.382..194N}. The detailed model can be written as follows:
\begin{align*}
	model_{Pexmon} = & phabs[1]*(zphabs[2]*powerlaw[3]+\\
	& constant[4] * powerlaw[5] +  mekal[6] +\\
	& pexmon[7]).
\end{align*}
In this model, the $phabs[1]$ component stands for the Galactic absorption is fixed at a column density of $5.16\times10^{20} cm^{-2}$, $powerlaw[3]$ represents the primary X-ray emission of the AGN, which is absorbed by obscured material ($zphabs[2]$). The primary X-ray emission is scattered by ionized gas in the polar regions, while the scattering component is usually little or not absorbed. The soft scattering component into our LOS is represented by $constant[4] * powerlaw[5]$. Thus the parameters of $powerlaw[5]$ are tied to the parameters of $powerlaw[3]$. The diffuse thermal radiation is denoted by $mekal[6]$, similar to Section~\ref{sec:basic}. We use $pexmon[7]$ to represent the reprocessed emission.

The \textit{Pexmon} model assumes a slab obscurer/reflector with an infinite optical depth, as may be expected in a standard geometrically thin accretion disc. The incident source in X-rays is the primary emission ($powerlaw$) from a hot electron corona.
Therefore, the photon index and normalization of $pexmon[7]$ are tied to the identical parameters of $powerlaw[3]$. The cutoff energy of $pexmon[7]$ is fixed as 500 keV. 
We are not able to constrain the Fe abundance (tied to the elemental abundance), so we fix it to its best-fit value.
The reflection fraction was fixed to -1 in the \textit{Pexmon} fit.

\begin{figure}
	\includegraphics[width=1.0\linewidth]{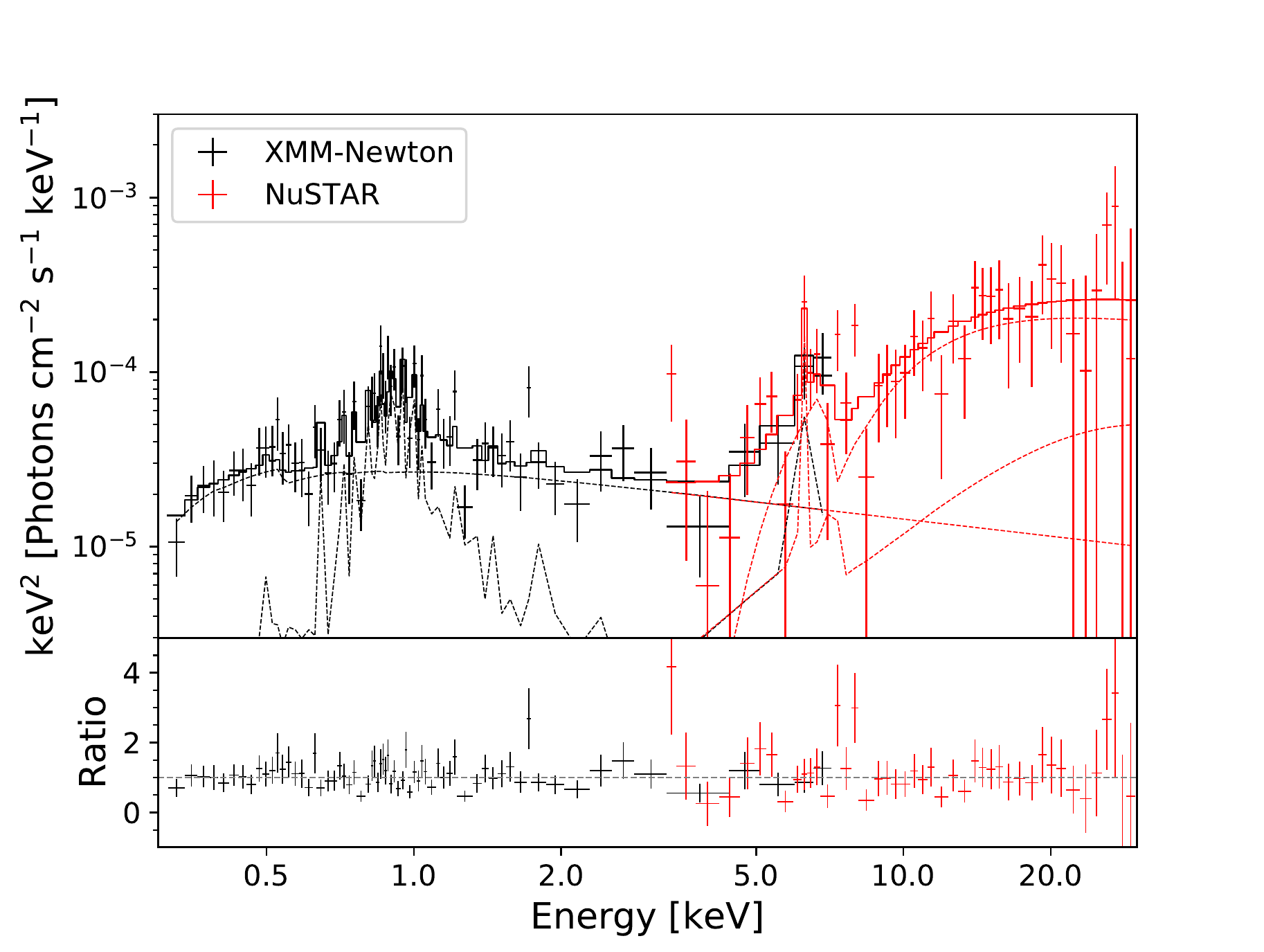}
	\caption{Pexmon model fit to the XMM-Newton PN (black) and NuSTAR (red) data. The top panel represents the best-fitting. The bottom panel shows the residuals.}
	\label{fig:pexmon}
\end{figure}

This model fits the data well ($\chi^2$/dof=101.38/105). The best-fitting and residuals are shown in Figure~\ref{fig:pexmon}. 
The \textit{Pexmon} model implies that the X-ray emission of central AGN is absorbed by Compton-thick material with $N_{\rm H} = (1.27\pm0.26)\times10^{24}\ \rm{cm^{-2}}$. The intrinsic 2--10~keV luminosity is $\rm{L}_{2-10}=(0.64\pm0.16)\times 10^{42}\ \rm{erg\ s^{-1}}$. The photon index is 2.44, suggesting that the spectrum of central AGN is very soft. 
We note that the scattering component is found to be $f_{\rm{scat}}= (3.22\pm 1.57)\%$ of the primary emission, and that the 2--10~keV luminosity of scattering component is $(2.05\pm 1.09)\times10^{40}\ \rm{erg\ s^{-1}}$. The diffuse thermal gas model ($mekal[6]$) adequately describes the soft X-rays with a temperature kT = 0.82~keV. More best-fit parameters for the model are listed in Table~\ref{Tab:pexmon-borus}.

\begin{deluxetable}{lcc}
	\label{Tab:pexmon-borus}
	\tablecaption{Best-fit parameters for \textit{Pexmon} and \textit{Borus}.}
	\tablewidth{0pt}
	\tablehead{
		\colhead{Parameter} & \colhead{\textit{Pexmon}} & \colhead{\textit{Borus}} 
	}
	\decimalcolnumbers
	\startdata
	$N_{\rm H,LOS}\ (10^{24}\ {\rm cm}^{-2})$ & $1.27\pm 0.26$&$0.97\pm 0.16$\\
	$\log N_{\rm H,Torus}\ ({\rm cm}^{-2})$ & $\cdots$&$24.09\pm 0.13$\\
	$\Gamma$ & $2.44\pm 0.13$&$2.38\pm 0.10$\\
	$f_{\rm{scat}}(\%)$ & $3.22\pm 1.52$&$2.50\pm 1.04$\\
	$f_{2-10,\rm{scat}}(10^{-14}\ \rm{erg\ s^{-1}\ cm^{-2}})$&$3.81\pm 2.04$& $4.28\pm 2.10$\\
	$L_{2-10,\rm{scat}}(10^{40}\ \rm{erg\ s^{-1}})$& $2.05\pm 1.09$ & $2.30\pm 1.13$\\
	kT(keV)& $0.82\pm 0.06$&$0.82\pm 0.06$\\
	$\theta_{\rm{torus}}({}^{\circ})$& \dots &25.0\\
	$\theta_{\rm{inc}}({}^{\circ})$& $76.23\pm 9.44$&84.4\\
	$f_{2-10,\rm{abs}}(10^{-12}\ \rm{erg\ s^{-1}\ cm^{-2}})$&$0.104^{+0.002}_{-0.034}$& $0.104^{+0.005}_{-0.018}$\\
	$L_{2-10,\rm{abs}}(10^{42}\ \rm{erg\ s^{-1}})$& $0.063^{+0.002}_{-0.013}$ & $0.063^{+0.005}_{-0.013}$\\
	$f_{2-10,\rm{int}}(10^{-12}\ \rm{erg\ s^{-1}\ cm^{-2}})$&$1.18\pm 0.30$& $1.72\pm 0.44$\\
	$L_{2-10,\rm{int}}(10^{42}\ \rm{erg\ s^{-1}})$& $0.64\pm 0.16$ & $0.93\pm 0.24$\\
	\hline\noalign{\smallskip}
	$\chi^2$/dof&101.38/105&101.35/105\\
	\enddata
	\tablecomments{$N_{\rm H,LOS}$: the column density of LOS. $N_{\rm H,Torus}$ : the average column density of torus. $\Gamma$: the photon index of the X-ray spectrum. $f_{\rm{scat}}$: the fraction of the scattering component. $f_{2-10,\rm{scat}}$: the flux density of the scattering component. $L_{2-10,\rm{scat}}$: the luminosity of the scattering component. kT: the temperature of diffuse thermal gas. $\theta_{\rm{torus}}$: the half-opening angle of torus. $\theta_{\rm{inc}}$: the inclination angle. $f_{2-10,\rm{abs}}$: the absorbed flux density. $L_{2-10,\rm{abs}}$: the absorbed luminosity. $f_{2-10,\rm{int}}$: the intrinsic flux density. $L_{2-10,\rm{int}}$: the intrinsic luminosity.}
\end{deluxetable}

\subsubsection{Borus model}\label{sec:borus}
Similarly, according to the analysis in Section~\ref{sec:basic}, we refit the spectrum of the reprocessed emission using the \textit{Borus} model \citep{2018ApJ...854...42B}. 
\textit{Borus} model is defined with the following command sequence in Xspec:
\begin{align*}
	model_{Borus} &=  phabs[1]*(zphabs[2]*cabs[3]*\\
	& powerlaw[4] + constant[5] * powerlaw[6]   \\
	& + mekal[7] + Borus02[8]).
\end{align*}
The $phabs[1]$, $constant[5] * powerlaw[6]$, and $mekal[7]$ components are equivalent to the ones in the Pexmon fit. The $zphabs[2]*cabs[3]$ represents LOS absorption at the redshift of the X-ray source (generally independent from the average column density of the torus), including Compton scattering losses out of the line of sight. The photon index and normalization of $Borus02[8]$ are also tied to the $powerlaw[3]$, and the cutoff energy is fixed as 500~keV. 
Some parameters can not be constrained, so we fix them to their best-fit values, i.e., half-opening angle, inclination angle, and iron abundance.

\begin{figure}
	\includegraphics[width=1.0\linewidth]{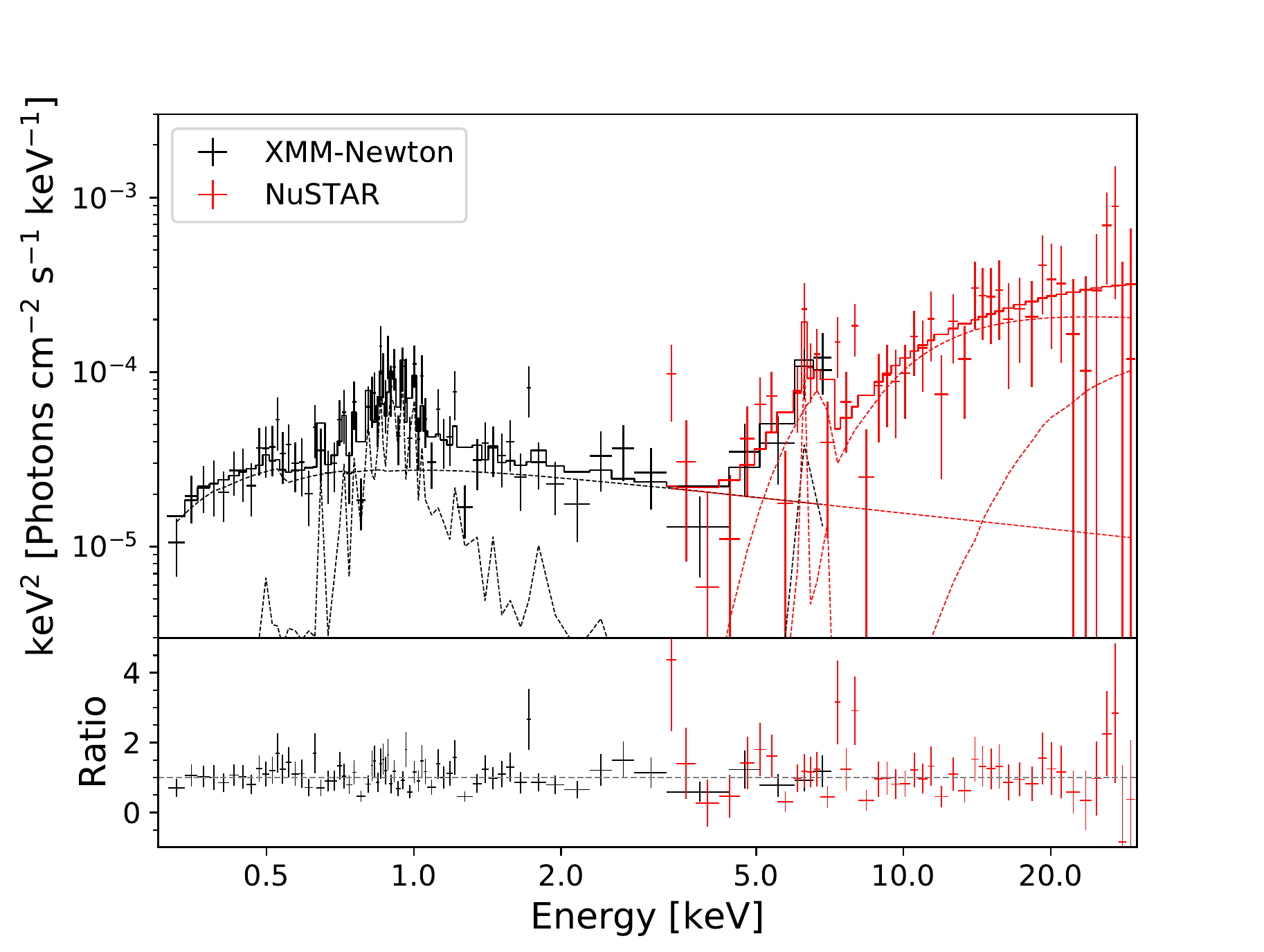}
	\caption{Borus model fit to the XMM-Newton PN (black) and NuSTAR (red) data. The top panel represents the best-fitting. The bottom panel shows the residuals.}
	\label{fig:borus}
\end{figure}
This model fits the data well ($\chi^2$/dof=101.35/105). The best-fitting and residuals are shown in Figure~\ref{fig:borus}. 
The column density of LOS is $N_{\rm H} = (0.97\pm 0.16)\times10^{24}\ \rm{cm^{-2}}$ , and the average column density of the torus is about $N_{\rm H,Torus} = 1.23\times10^{24}\ \rm{cm^{-2}}$. 
The intrinsic 2--10~keV luminosity is $\rm{L}_{2-10}=(0.93\pm 0.24)\times 10^{42}\ \rm{erg\ s^{-1}}$.
More best-fit parameters for the model are listed in Table~\ref{Tab:pexmon-borus}.

Through the above analysis and the broadband X-ray spectrum fitting of two physical models (\textit{Pexmon} and \textit{Borus}), we have obtained some parameters about the AGN in NGC 449. Most of the parameters obtained by the two models are the identical within the error range, i.e., column density of LOS, the photon index, the luminosity of scattering component, the temperature of diffuse thermal gas, and the absorbed luminosity. These suggest that these parameters can be well constrained in the X-ray band. However, the fraction of the scattering component and the intrinsic luminosity can not be well constrained. Therefore, we need information from other bands to constrain these parameters.

\subsection{Mid-IR spectrum Analysis}\label{sec:mid-IR-analysis}
In this section, we use DeblendIRS\footnote{For more details, refer to \url{http://www.denebola.org/ahc/deblendIRS/}} \citep{2015ApJ...803..109H} to analyze the mid-IR spectrum of NGC 449. DeblendIRS is an IDL package that fits the mid-IR spectra with a linear combination of three spectral templates, i.e., a “pure” AGN template, a “pure” stellar template, and a “pure” Polycyclic Aromatic Hydrocarbon (PAH, which accounts for the interstellar emission) template. 
The mid-IR spectra of the sources can be well decomposed into the contributions of different components using this package. Thus we obtain some physical properties of the AGNs and their host galaxies, such as the silicate strength\footnote{The silicate strength is defined as $$\rm{S}_{\rm{Si}}=\ln \frac{F(\lambda_p)}{F_C(\lambda_p)},$$ where ${\rm F}(\lambda_p)$and ${\rm F}_C(\lambda_p)$ stand for the maximum flux density of the silicate line profile near 9.7~\micron\ and the corresponding flux density of the underlying continuum profile,respectively. $\rm{S}_{\rm{Si}}$ is a negative value, which means that the central radiation of the AGN is absorbed by the torus.The optical depth of the silicate absorption $\tau_{9.7}= -\rm{S}_{\rm{Si}}$. } ($\rm{S}_{\rm{Si}}$),  spectral index ($\alpha$),  AGN emission at rest-frame 6~\micron, and 12~\micron.

\begin{figure}
	\includegraphics[width=1.0\linewidth]{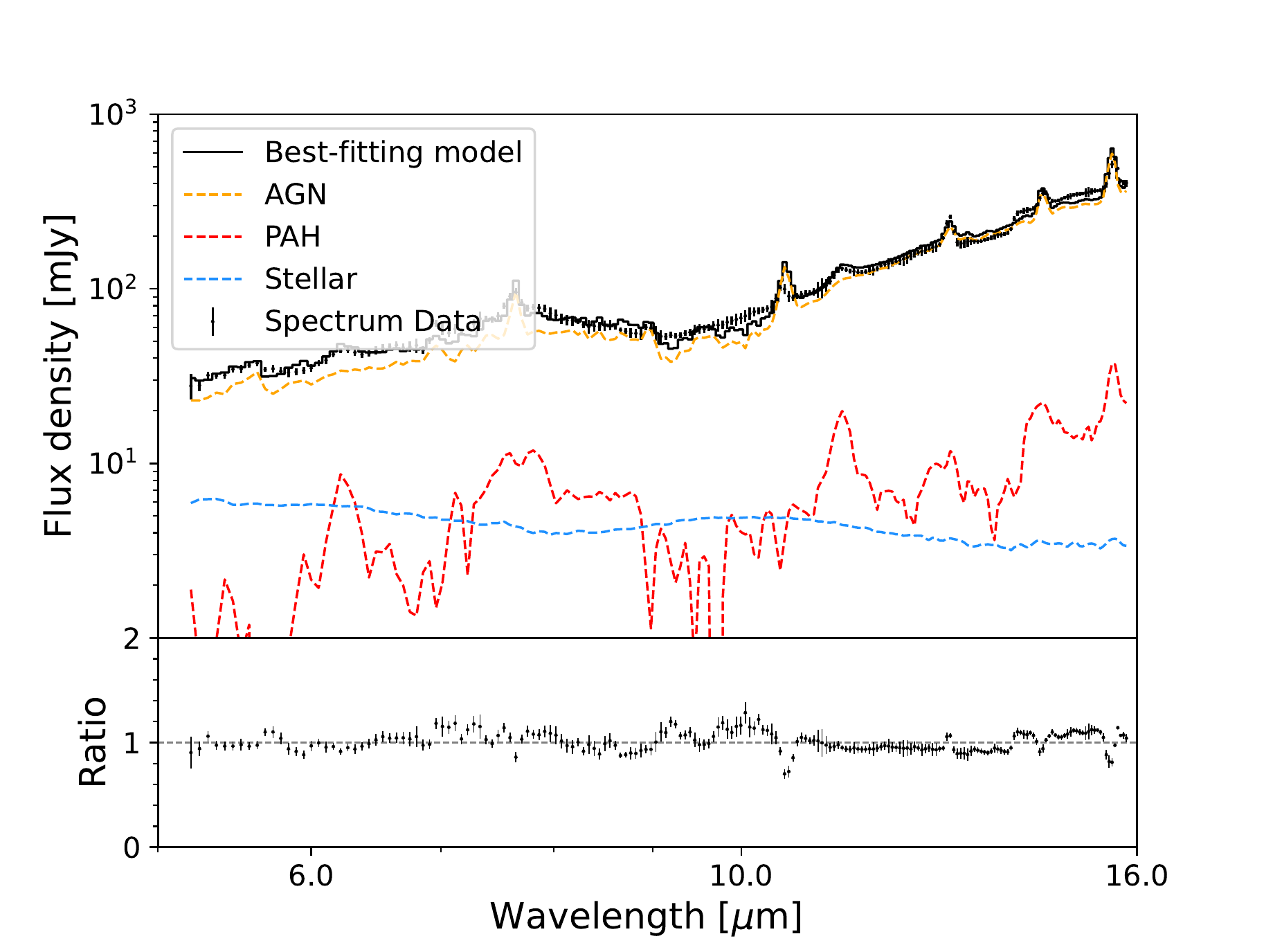}
	\caption{The best-fitting of the mid-IR spectrum. The black line indicates the best-fit model. The blue, red, and orange dash lines represent stellar, PAH, and AGN emission, respectively. The bottom panel shows the residuals.}
	\label{fig:mid-spectrum}
\end{figure}
We use this package to fit the mid-IR spectrum (5 -- 16~\micron) and provide the best-fitting results. Figure~\ref{fig:mid-spectrum} shows the best-fitting of the mid-IR spectrum of NGC 449.  
From this fitting, we derive a silicate strength of $-0.71$, a luminosity at 6~\micron\ of $7.65\times 10^{42} \rm{erg\ s^{-1}}$, and a luminosity at 12~\micron\ of $1.84\times 10^{43} \rm{erg\ s^{-1}}$.

\subsection{Analysis of Multi-band SED} \label{sec:SED-fitting}
In this section, we analyze the Multi-band SED of this source using Code Investigating GALaxy Emission \citep[CIGALE 2020.0, ][]{2019A&A...622A.103B}. CIGALE 2020.0 is an open python code designed to estimate the physical properties (i.e., SFR, stellar mass, AGN luminosity) of galaxies and AGNs. 
We fit the SED of NGC 449 with the same modules and parameters as \cite{2020MNRAS.492.1887G}. For details, please refer to Section 3 of \cite{2020MNRAS.492.1887G}. 
\begin{figure}
	\includegraphics[width=1.0\linewidth]{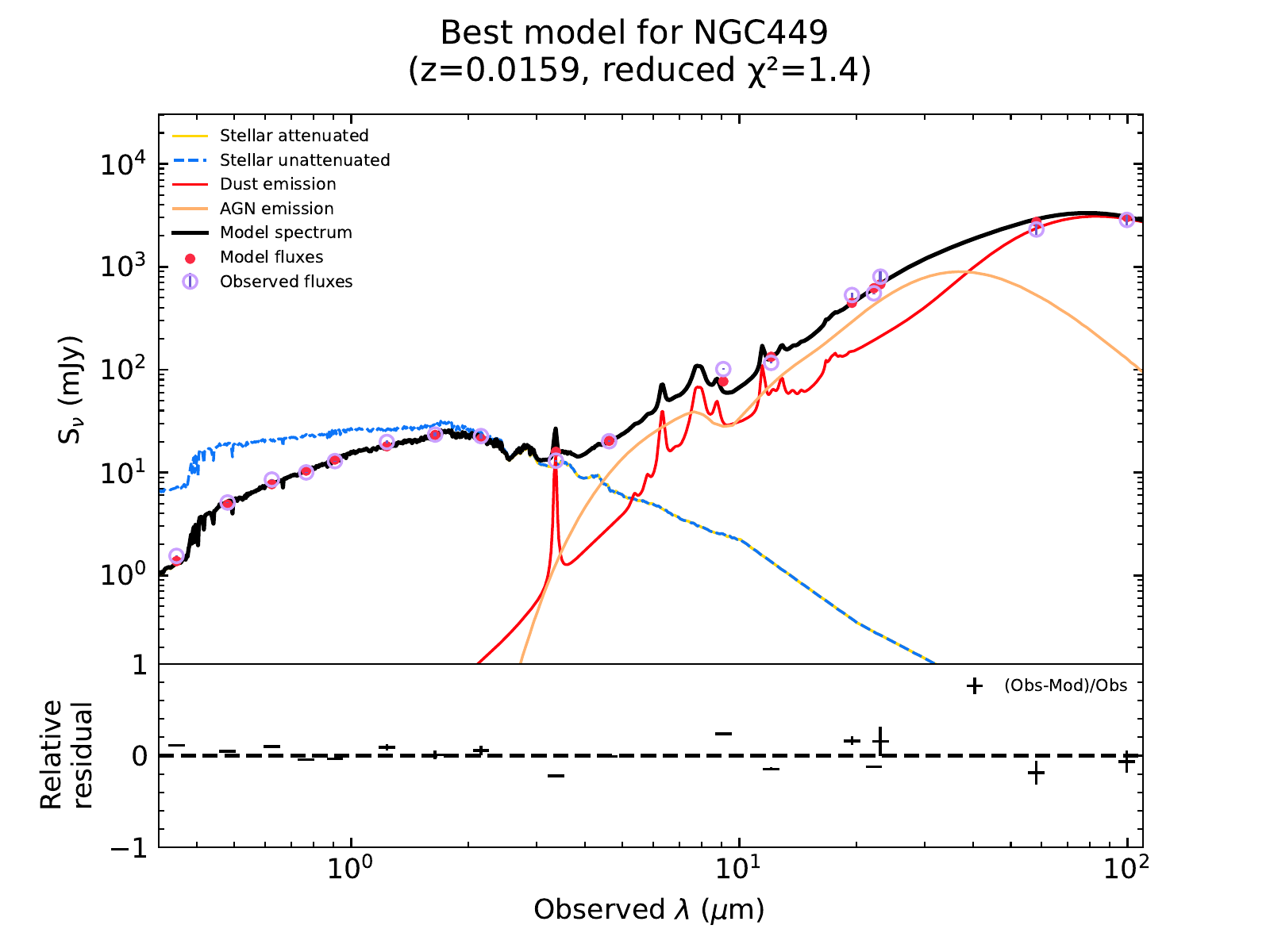}
	\caption{The best-fit SED for NGC 449. The black line indicates the best-fit model. The blue, yellow, red, and orange lines represent unattenuated stellar, attenuated stellar, dust, and AGN emission, respectively. The lower panel indicates residual of the best fitting.}
	\label{fig:SED}
\end{figure}

Figure~\ref{fig:SED} presents the best-fit SED for this source.
In addition, we also obtain some physical parameters of NGC 449 by the SED fitting, such as SFR, stellar mass, and AGN luminosity. Table~\ref{Tab:SED-parameters} lists several parameters that may be used in this work. The SFR estimate by the SED fitting is more dependent on the star formation history model, so we also use the calibration from \cite{2005ApJ...625...23B} to estimate SFR by the UV and IR luminosities, scaled to \cite{2003PASP..115..763C} initial mass function:
\begin{equation}
	\rm{SFR}(\rm{M_\odot/yr}) = 1.09\times 10^{-9} (3.3 \rm{L_{UV}}+\rm{L_{IR}})
\end{equation}
where $\rm{L_{UV}}= \nu L_\nu$ is an estimation of the integrated 1216 -- 3000\AA \ rest-frame UV luminosity, 
and $\rm{L_{IR}}$ is the 8 -- 1000~\micron\ rest-frame IR luminosity. Both $\rm{L_{UV}}$  and $\rm{L_{IR}}$ (listed in Table~\ref{Tab:SED-parameters}) are in units of L$_\odot$.  The SFR of NGC 449 is 3.12 M$_\odot$/yr which is estimated by the UV and IR luminosities.

\begin{deluxetable}{lc}
	\label{Tab:SED-parameters}
	\tablecaption{Some parameters are obtained/derived by the SED fitting.}
	\tablewidth{0pt}
	\tablehead{
		\colhead{Parameter } & \colhead{Value} 
	}
	\decimalcolnumbers
	\startdata
	M$_*$ (M$_\odot$) &$(5.28\pm 0.29)\times 10^9$ \\
	SFR$_{\rm{SED}}$ (M$_\odot$ yr$^{-1}$) & $2.02\pm 0.25$\\
	AGN Luminosity ($\rm{erg\ s^{-1}}$)&$(5.35\pm 0.58)\times 10^{43}$ \\
	$\chi^2$/dof & 1.40\\
	\hline
	$\rm{L_{UV}}$( L$_\odot$)&$8.22\times 10^{8}$ \\
	$\rm{L_{IR}}$( L$_\odot$)&$2.61\times 10^{10}$ \\
	SFR$_{\rm{UV}+\rm{IR}}$  (M$_\odot$ yr$^{-1}$) & 3.12\\
	\enddata
\end{deluxetable}

\section{Results and Discussion}\label{sec:result}

We have obtained some properties of NGC 449 through individual band analyses. However, some properties are not well-constrained. Combined with the multiband results, the properties of this source will be better constrained. In addition, we will also derive some other properties of this source.

\subsection{The properties of the AGN central engine}

The AGN at the center of NGC 449 is an optical type-2 AGN, so we cannot directly observe its central emission (e.g., accretion disk, corona). It is difficultly to understand the coronal (intrinsic) emission only by the X-ray spectrum fitting. For heavily obscured AGNs, the estimation of intrinsic luminosities (flux densities) is very dependent on the X-ray model employed. Although Table~\ref{Tab:pexmon-borus} presents the intrinsic 2--10 keV luminosities (flux densities) obtained by the \textit{Pexmon} and \textit{Borus} models, they are unreliable.
To derive a reliable intrinsic luminosity, we had better estimate it in other ways.
Fortunately,  there is a very close correlation between 2--10 keV and mid-IR 12~\micron \ luminosities of AGNs \citep[e.g.,][]{2015MNRAS.454..766A}. 
In Section~\ref{sec:mid-IR-analysis}, we have obtained the mid-IR 12~\micron \ luminosity ($\nu \rm{L}_{\nu}(12~\micron) = 1.84\times 10^{43} \rm{erg\ s^{-1}}$) of central AGN.
We use Equation 2 of \cite{2015MNRAS.454..766A} to estimate intrinsic 2--10 keV luminosity of $(8.54\pm0.75)\times 10^{42}\ \rm{erg\  s^{-1}}$.
Moreover, the fraction of the scattering component is also not well constrained.
However, the \textit{Pexmon} and \textit{Borus} models have well-constrained the scattering luminosity, whose average value is $(2.18\pm 1.11)\times10^{40}\ \rm{erg\ s^{-1}}$.  
The relationship between the intrinsic luminosity and scattering luminosity is $f_{\rm{scat}}={\rm L_{scat}}/{\rm L_{Int}}$.
So the scattering fraction of the X-ray emission of central AGN is about $(0.26\pm 0.13)\%$.

Since the UV-to-optical continuum of this source is mainly contributed by the host galaxy, we cannot obtain the bolometric emission of the accretion disk by SED fitting. However, the coronal and torus emissions can well track the thermal radiation of the accretion disk. So they can be used to estimate the bolometric luminosity of type-2 AGNs. Using the intrinsic 2--10 keV luminosity, we solve the Equation 21 of \cite{2004MNRAS.351..169M} to estimate the bolometric luminosity $\rm{L_{bol}} = 1.44\times 10^{44}\ \rm{erg\ s^{-1}}$ (bolometric correction $k_{bol}\simeq 16.85$).  This bolometric luminosity is consistent with that provided by \cite{2002ApJ...579..530W}.

The mass accretion rate is a measurement of SMBH mass growth and is related to bolometric luminosity.
The relationship is written as $\dot{m}_{\rm{acc}}=\rm{L_{bol}} /\eta c^2$, where $\eta$ is the efficiency that converts the rest-mass energy of accreted material into radiation. Assuming $\eta = 0.1$ \citep{2002apa..book.....F}, we estimate the mass accretion rate of NGC449, which is $2.54 \times 10^{-2} \rm{M}_\odot\ \rm{yr}^{-1}$. 
The Eddington ratio is a measurement of the SMBH accretion efficiency and is defined as $\lambda_{\rm{Edd}} = \rm{L_{bol}/L_{Edd}}$, where $\rm{L_{bol}}$ is the bolometric luminosity of the AGN, $\rm{L_{Edd}}$ is the Eddington luminosity of the AGN ($1.3\times10^{38}\ \rm{M_{BH}/M_\odot\ erg\ s^{-1}}$). 
Using the SMBH mass ($1.32\times 10^7\ \rm{M}_\odot$) given in \cite{2002ApJ...579..530W}, we estimate the Eddington Luminosity to be $\rm{L_{Edd}} \simeq 1.72\times10^{45}\ \rm{erg\ s^{-1}}$.
The Eddington ratio is about $\lambda_{\rm{Edd}} =8.39\times 10^{-2}$.
The Eddington ratio of most CT-AGNs in the local universe is similar to our result \citep[e.g.,][]{2022ApJS..260...30T}. 
\subsection{The properties of the torus}
Previous studies have attempted to constrain the column density of NGC449 using XMM-Newton data or the ratio of X-ray to high-ionized emission lines ([O III]5007, [Ne V]3426) luminosities \citep{2005A&A...444..119G,2005ApJ...634..161H,2010A&A...519A..92G}. Those results suggested that the center of NGC449 hosted a CT-AGN.
We use the latest X-ray data above 10 keV to construct a broadband X-ray spectrum.
Compared with the previous studies, we can better constrain the parameters of the torus in Section~\ref{sec:X-ray-analysis}.
In the \textit{Pexmon} model, the column density of the torus is obtained as $(1.27\pm 0.26)\times 10^{24}\ \rm{cm}^{-2}$, indicating that it may be a CT-AGN. In Section~\ref{sec:borus}, we re-fit the broadband X-ray spectrum using the \textit{Borus} model and obtain a series of parameters about torus. Among these parameters, the column density of LOS is $(0.97\pm 0.16)\times 10^{24}\ \rm{cm}^{-2}$, and the average column density of the torus is about $1.23\times 10^{24}\ \rm{cm}^{-2}$, which is close to Compton-thick. 
In conclusion, the column density of this source we obtain is similar to that estimated by previous works \citep{2005A&A...444..119G,2005ApJ...634..161H,2010A&A...519A..92G}, but it is well-constrained.

In Section~\ref{sec:mid-IR-analysis}, we decomposed the contribution of AGN by the mid-IR spectrum fitting, and derived the silicate strength of the AGN component.
Using the silicate strength, we can derive the optical depth of silicate absorption as $\tau_{9.7}=0.71$. The optical depth of silicate absorption is a good way to track the dust in the torus, i.e., the V-band extinction is $A_{\rm V} =19 \times A_{9.7}= 19 \times 1.086\tau_{9.7}=14.65\ \rm{mag}$ \citep{1985MNRAS.215..425R}. Therefore, the gas-to-dust ratio of the torus is $N_{\rm H}/A_{\rm V}=8.40\times 10^{22}\ \rm{cm}^{-2}\ \rm{mag}^{-1}$ ($N_{\rm H}$ is the average column density of torus).
This gas-to-dust ratio is about 45 times that of the Galaxy \citep[$N_{\rm H}/A_{\rm V}=1.87\times 10^{21}\ \rm{cm}^{-2}\ \rm{mag}^{-1}$,][]{2003ARA&A..41..241D}.
Even this gas-to-dust ratio is higher than that of nearby CT-AGN in the Circinus galaxy \citep{2019ApJ...877...95T}, which is the most obscured AGN in the nearby universe. 
Such a high gas-to-dust ratio of this source means that the radiation of the central AGN may have destroyed dust in the torus.

\subsection{The evolutionary stage of the source}
Many studies have suggested a co-evolution scheme between host galaxies and AGNs \citep[e.g.,][]{2013ARA&A..51..511K}, such as $\rm{M_{BH}}-\sigma_*$ relation \citep[e.g.,][]{2000ApJ...539L...9F}, AGN feedback \citep[e.g.,][]{2006MNRAS.370..645B}. Some studies have suggested that the host galaxies of heavily obscured AGNs appear to be at the stage of intense star formation \citep[e.g.,][]{2006ApJS..163....1H}. These intense star-forming activities may be associated with AGNs, i.e. AGNs trigger star formation of host galaxies.

NGC 449 is a late-type galaxy, meaning it is in an early stage of galaxy evolution. 
Its center hosts a heavily obscured AGN, which is considered to be at the early stage of AGN evolution. These suggest that the host galaxy and central AGN of NGC 449 are at the early stage of co-evolution.
In Section~\ref{sec:SED-fitting}, we derive some parameters about the host galaxy, including SFR and stellar mass. The specific star formation rate ($\rm{sSFR = SFR}/ \rm{M}_*$) of this source is approximately $0.59\ \rm{Gyr}^{-1}$. This value is consistent with other star formation galaxies at similar redshifts  \citep{2007ApJS..173..267S}. This result suggests that the central AGN does not appear to trigger intense star formation in its host galaxy.
\section{Summary}\label{sec:summary}
We studied a heavily obscured AGN in NGC 449 using the multiwavelength data, which included NuSTAR, XMM-Newton,  Spitzer, and multi-band photometric data.
After, we analyzed its X-ray spectrum and obtained the column density ($N_{\rm H}\simeq 10^{24}  \rm{cm}^{-2}$), the photon index ($\Gamma \simeq 2.4$), the luminosity (flux density) of the scattering component, and the temperature ($kT \simeq 0.82~\rm{keV}$) of diffuse thermal gas. However, some parameters could not be well constrained, i.e., the fraction of the scattering component, the intrinsic flux density, and luminosity. We fitted its mid-IR spectrum and decomposed the AGN contribution to derive some AGN parameters, and used its SED to obtain some parameters.
Combining the information from the mid-IR spectrum and SED, we first derived the intrinsic X-ray luminosity ($\simeq 8.54\times 10^{42} \ \rm{erg\ s}^{-1}$) and the scattering fraction of primary X-ray emission ($\simeq 0.26\%$).
In addition, we also derived the following results:
\begin{enumerate}
	\item The bolometric luminosity of the AGN is about $1.44\times 10^{44} \ \rm{erg\ s}^{-1}$. The mass accretion rate of central AGN  is about $2.54 \times 10^{-2} \rm{M}_\odot\ \rm{yr}^{-1}$, and the Eddington ratio is $8.39\times 10^{-2}$. 
	\item The torus of this AGN has a high gas-to-dust ratio ($N_{\rm H}/A_{\rm V} =8.40\times 10^{22}\ \rm{cm}^{-2}\ \rm{mag}^{-1}$), which is about 45 times that of the Galaxy. Such a high gas-to-dust ratio means that the radiation of the central AGN may have destroyed dust in the torus.
	\item The host galaxy and the central AGN are both in the early stage of co-evolution.
\end{enumerate}

\begin{acknowledgments}
	We sincerely thank the anonymous referee for useful suggestions.
We acknowledge financial support from the National Key Research and Development Program of China grant (No. 2017YFA0402703) and National Natural Science Foundation of China grant (NSFC; grant Nos.11733002). We acknowledge the science research grants from the China Manned Space Project with NO. CMS-CSST-2021-A07. 
Yongyun Chen is gratefulfor financial support from the National Natural Science Foundation of China (No. 12203028). 
This work is supported by the youth project of Yunnan Provincial Science and Technology Department (202101AU070146, 2103010006). 
Yongyun Chen is grateful for funding for the training Program for talents in Xingdian, Yunnan Province.
Nan Ding thanks for the financial support from the National Natural Science Foundation of China (No. 12103022) and the Special Basic Cooperative Research Programs of Yunnan Provincial Undergraduate Universities'Association (No. 202101BA070001-043).
\end{acknowledgments}

%

\vspace{5mm}


\software{ numpy, pandas,  matplotlib \citep{2007CSE.....9...90H}, astropy\citep{2013A&A...558A..33A}
          }





\bibliography{bibtex}{}

\begin{thebibliography}{}
\expandafter\ifx\csname natexlab\endcsname\relax\def\natexlab#1{#1}\fi
\providecommand{\url}[1]{\href{#1}{#1}}
\providecommand{\dodoi}[1]{doi:~\href{http://doi.org/#1}{\nolinkurl{#1}}}
\providecommand{\doeprint}[1]{\href{http://ascl.net/#1}{\nolinkurl{http://ascl.net/#1}}}
\providecommand{\doarXiv}[1]{\href{https://arxiv.org/abs/#1}{\nolinkurl{https://arxiv.org/abs/#1}}}

\bibitem[{{Abazajian} {et~al.}(2009){Abazajian}, {Adelman-McCarthy},
  {Ag{\"u}eros}, {Allam}, {Allende Prieto}, {An}, {Anderson}, {Anderson},
  {Annis}, {Bahcall}, {Bailer-Jones}, {Barentine}, {Bassett}, {Becker},
  {Beers}, {Bell}, {Belokurov}, {Berlind}, {Berman}, {Bernardi}, {Bickerton},
  {Bizyaev}, {Blakeslee}, {Blanton}, {Bochanski}, {Boroski}, {Brewington},
  {Brinchmann}, {Brinkmann}, {Brunner}, {Budav{\'a}ri}, {Carey}, {Carliles},
  {Carr}, {Castander}, {Cinabro}, {Connolly}, {Csabai}, {Cunha}, {Czarapata},
  {Davenport}, {de Haas}, {Dilday}, {Doi}, {Eisenstein}, {Evans}, {Evans},
  {Fan}, {Friedman}, {Frieman}, {Fukugita}, {G{\"a}nsicke}, {Gates},
  {Gillespie}, {Gilmore}, {Gonzalez}, {Gonzalez}, {Grebel}, {Gunn},
  {Gy{\"o}ry}, {Hall}, {Harding}, {Harris}, {Harvanek}, {Hawley}, {Hayes},
  {Heckman}, {Hendry}, {Hennessy}, {Hindsley}, {Hoblitt}, {Hogan}, {Hogg},
  {Holtzman}, {Hyde}, {Ichikawa}, {Ichikawa}, {Im}, {Ivezi{\'c}}, {Jester},
  {Jiang}, {Johnson}, {Jorgensen}, {Juri{\'c}}, {Kent}, {Kessler}, {Kleinman},
  {Knapp}, {Konishi}, {Kron}, {Krzesinski}, {Kuropatkin}, {Lampeitl},
  {Lebedeva}, {Lee}, {Lee}, {French Leger}, {L{\'e}pine}, {Li}, {Lima}, {Lin},
  {Long}, {Loomis}, {Loveday}, {Lupton}, {Magnier}, {Malanushenko},
  {Malanushenko}, {Mandelbaum}, {Margon}, {Marriner}, {Mart{\'\i}nez-Delgado},
  {Matsubara}, {McGehee}, {McKay}, {Meiksin}, {Morrison}, {Mullally}, {Munn},
  {Murphy}, {Nash}, {Nebot}, {Neilsen}, {Newberg}, {Newman}, {Nichol},
  {Nicinski}, {Nieto-Santisteban}, {Nitta}, {Okamura}, {Oravetz}, {Ostriker},
  {Owen}, {Padmanabhan}, {Pan}, {Park}, {Pauls}, {Peoples}, {Percival}, {Pier},
  {Pope}, {Pourbaix}, {Price}, {Purger}, {Quinn}, {Raddick}, {Re Fiorentin},
  {Richards}, {Richmond}, {Riess}, {Rix}, {Rockosi}, {Sako}, {Schlegel},
  {Schneider}, {Scholz}, {Schreiber}, {Schwope}, {Seljak}, {Sesar}, {Sheldon},
  {Shimasaku}, {Sibley}, {Simmons}, {Sivarani}, {Allyn Smith}, {Smith},
  {Smol{\v{c}}i{\'c}}, {Snedden}, {Stebbins}, {Steinmetz}, {Stoughton},
  {Strauss}, {SubbaRao}, {Suto}, {Szalay}, {Szapudi}, {Szkody}, {Tanaka},
  {Tegmark}, {Teodoro}, {Thakar}, {Tremonti}, {Tucker}, {Uomoto}, {Vanden
  Berk}, {Vandenberg}, {Vidrih}, {Vogeley}, {Voges}, {Vogt}, {Wadadekar},
  {Watters}, {Weinberg}, {West}, {White}, {Wilhite}, {Wonders}, {Yanny},
  {Yocum}, {York}, {Zehavi}, {Zibetti}, \& {Zucker}}]{2009ApJS..182..543A}
{Abazajian}, K.~N., {Adelman-McCarthy}, J.~K., {Ag{\"u}eros}, M.~A., {et~al.}
  2009, \apjs, 182, 543, \dodoi{10.1088/0067-0049/182/2/543}

\bibitem[{{Ajello} {et~al.}(2008){Ajello}, {Greiner}, {Sato}, {Willis},
  {Kanbach}, {Strong}, {Diehl}, {Hasinger}, {Gehrels}, {Markwardt}, \&
  {Tueller}}]{2008ApJ...689..666A}
{Ajello}, M., {Greiner}, J., {Sato}, G., {et~al.} 2008, \apj, 689, 666,
  \dodoi{10.1086/592595}

\bibitem[{{Antonucci}(1993)}]{1993ARA&A..31..473A}
{Antonucci}, R. 1993, \araa, 31, 473,
  \dodoi{10.1146/annurev.aa.31.090193.002353}

\bibitem[{{Arnaud}(1996)}]{1996ASPC..101...17A}
{Arnaud}, K.~A. 1996, in Astronomical Society of the Pacific Conference Series,
  Vol. 101, Astronomical Data Analysis Software and Systems V, ed. G.~H.
  {Jacoby} \& J.~{Barnes}, 17

\bibitem[{{Asmus} {et~al.}(2015){Asmus}, {Gandhi}, {H{\"o}nig}, {Smette}, \&
  {Duschl}}]{2015MNRAS.454..766A}
{Asmus}, D., {Gandhi}, P., {H{\"o}nig}, S.~F., {Smette}, A., \& {Duschl}, W.~J.
  2015, \mnras, 454, 766, \dodoi{10.1093/mnras/stv1950}

\bibitem[{{Astropy Collaboration} {et~al.}(2013){Astropy Collaboration},
  {Robitaille}, {Tollerud}, {Greenfield}, {Droettboom}, {Bray}, {Aldcroft},
  {Davis}, {Ginsburg}, {Price-Whelan}, {Kerzendorf}, {Conley}, {Crighton},
  {Barbary}, {Muna}, {Ferguson}, {Grollier}, {Parikh}, {Nair}, {Unther},
  {Deil}, {Woillez}, {Conseil}, {Kramer}, {Turner}, {Singer}, {Fox}, {Weaver},
  {Zabalza}, {Edwards}, {Azalee Bostroem}, {Burke}, {Casey}, {Crawford},
  {Dencheva}, {Ely}, {Jenness}, {Labrie}, {Lim}, {Pierfederici}, {Pontzen},
  {Ptak}, {Refsdal}, {Servillat}, \& {Streicher}}]{2013A&A...558A..33A}
{Astropy Collaboration}, {Robitaille}, T.~P., {Tollerud}, E.~J., {et~al.} 2013,
  \aap, 558, A33, \dodoi{10.1051/0004-6361/201322068}

\bibitem[{{Balokovi{\'c}} {et~al.}(2018){Balokovi{\'c}}, {Brightman},
  {Harrison}, {Comastri}, {Ricci}, {Buchner}, {Gandhi}, {Farrah}, \&
  {Stern}}]{2018ApJ...854...42B}
{Balokovi{\'c}}, M., {Brightman}, M., {Harrison}, F.~A., {et~al.} 2018, \apj,
  854, 42, \dodoi{10.3847/1538-4357/aaa7eb}

\bibitem[{{Bell} {et~al.}(2005){Bell}, {Papovich}, {Wolf}, {Le Floc'h},
  {Caldwell}, {Barden}, {Egami}, {McIntosh}, {Meisenheimer},
  {P{\'e}rez-Gonz{\'a}lez}, {Rieke}, {Rieke}, {Rigby}, \&
  {Rix}}]{2005ApJ...625...23B}
{Bell}, E.~F., {Papovich}, C., {Wolf}, C., {et~al.} 2005, \apj, 625, 23,
  \dodoi{10.1086/429552}

\bibitem[{{Boorman} {et~al.}(2018){Boorman}, {Gandhi}, {Balokovi{\'c}},
  {Brightman}, {Harrison}, {Ricci}, \& {Stern}}]{2018MNRAS.477.3775B}
{Boorman}, P.~G., {Gandhi}, P., {Balokovi{\'c}}, M., {et~al.} 2018, \mnras,
  477, 3775, \dodoi{10.1093/mnras/sty861}

\bibitem[{{Boquien} {et~al.}(2019){Boquien}, {Burgarella}, {Roehlly}, {Buat},
  {Ciesla}, {Corre}, {Inoue}, \& {Salas}}]{2019A&A...622A.103B}
{Boquien}, M., {Burgarella}, D., {Roehlly}, Y., {et~al.} 2019, \aap, 622, A103,
  \dodoi{10.1051/0004-6361/201834156}

\bibitem[{{Bower} {et~al.}(2006){Bower}, {Benson}, {Malbon}, {Helly}, {Frenk},
  {Baugh}, {Cole}, \& {Lacey}}]{2006MNRAS.370..645B}
{Bower}, R.~G., {Benson}, A.~J., {Malbon}, R., {et~al.} 2006, \mnras, 370, 645,
  \dodoi{10.1111/j.1365-2966.2006.10519.x}

\bibitem[{{Cappi} {et~al.}(2006){Cappi}, {Panessa}, {Bassani}, {Dadina}, {Di
  Cocco}, {Comastri}, {della Ceca}, {Filippenko}, {Gianotti}, {Ho}, {Malaguti},
  {Mulchaey}, {Palumbo}, {Piconcelli}, {Sargent}, {Stephen}, {Trifoglio}, \&
  {Weaver}}]{2006A&A...446..459C}
{Cappi}, M., {Panessa}, F., {Bassani}, L., {et~al.} 2006, \aap, 446, 459,
  \dodoi{10.1051/0004-6361:20053893}

\bibitem[{{Chabrier}(2003)}]{2003PASP..115..763C}
{Chabrier}, G. 2003, \pasp, 115, 763, \dodoi{10.1086/376392}

\bibitem[{Comastri(2004)}]{2004ASSL..308..245C}
Comastri, A. 2004, Compton Thick AGN: the dark side of the X-ray background
  (Springer Netherlands), 245--272

\bibitem[{{Draine}(2003)}]{2003ARA&A..41..241D}
{Draine}, B.~T. 2003, \araa, 41, 241,
  \dodoi{10.1146/annurev.astro.41.011802.094840}

\bibitem[{{Ferrarese} \& {Merritt}(2000)}]{2000ApJ...539L...9F}
{Ferrarese}, L., \& {Merritt}, D. 2000, \apjl, 539, L9, \dodoi{10.1086/312838}

\bibitem[{{Frank} {et~al.}(2002){Frank}, {King}, \&
  {Raine}}]{2002apa..book.....F}
{Frank}, J., {King}, A., \& {Raine}, D.~J. 2002, {Accretion Power in
  Astrophysics: Third Edition}

\bibitem[{{Gabriel} {et~al.}(2004){Gabriel}, {Denby}, {Fyfe}, {Hoar}, {Ibarra},
  {Ojero}, {Osborne}, {Saxton}, {Lammers}, \& {Vacanti}}]{2004ASPC..314..759G}
{Gabriel}, C., {Denby}, M., {Fyfe}, D.~J., {et~al.} 2004, in Astronomical
  Society of the Pacific Conference Series, Vol. 314, Astronomical Data
  Analysis Software and Systems (ADASS) XIII, ed. F.~{Ochsenbein}, M.~G.
  {Allen}, \& D.~{Egret}, 759

\bibitem[{{Gandhi} {et~al.}(2017){Gandhi}, {Annuar}, {Lansbury}, {Stern},
  {Alexander}, {Bauer}, {Bianchi}, {Boggs}, {Boorman}, {Brandt}, {Brightman},
  {Christensen}, {Comastri}, {Craig}, {Del Moro}, {Elvis}, {Guainazzi},
  {Hailey}, {Harrison}, {Koss}, {Lamperti}, {Malaguti}, {Masini}, {Matt},
  {Puccetti}, {Ricci}, {Rivers}, {Walton}, \& {Zhang}}]{2017MNRAS.467.4606G}
{Gandhi}, P., {Annuar}, A., {Lansbury}, G.~B., {et~al.} 2017, \mnras, 467,
  4606, \dodoi{10.1093/mnras/stx357}

\bibitem[{{Gilli} {et~al.}(2010){Gilli}, {Vignali}, {Mignoli}, {Iwasawa},
  {Comastri}, \& {Zamorani}}]{2010A&A...519A..92G}
{Gilli}, R., {Vignali}, C., {Mignoli}, M., {et~al.} 2010, \aap, 519, A92,
  \dodoi{10.1051/0004-6361/201014039}

\bibitem[{{Goulding} {et~al.}(2011){Goulding}, {Alexander}, {Mullaney},
  {Gelbord}, {Hickox}, {Ward}, \& {Watson}}]{2011MNRAS.411.1231G}
{Goulding}, A.~D., {Alexander}, D.~M., {Mullaney}, J.~R., {et~al.} 2011,
  \mnras, 411, 1231, \dodoi{10.1111/j.1365-2966.2010.17755.x}

\bibitem[{{Guainazzi} {et~al.}(2005){Guainazzi}, {Matt}, \&
  {Perola}}]{2005A&A...444..119G}
{Guainazzi}, M., {Matt}, G., \& {Perola}, G.~C. 2005, \aap, 444, 119,
  \dodoi{10.1051/0004-6361:20053643}

\bibitem[{{Guo} {et~al.}(2020){Guo}, {Gu}, {Ding}, {Contini}, \&
  {Chen}}]{2020MNRAS.492.1887G}
{Guo}, X., {Gu}, Q., {Ding}, N., {Contini}, E., \& {Chen}, Y. 2020, \mnras,
  492, 1887, \dodoi{10.1093/mnras/stz3589}

\bibitem[{{Harrison} {et~al.}(2013){Harrison}, {Craig}, {Christensen},
  {Hailey}, {Zhang}, {Boggs}, {Stern}, {Cook}, {Forster}, {Giommi},
  {Grefenstette}, {Kim}, {Kitaguchi}, {Koglin}, {Madsen}, {Mao}, {Miyasaka},
  {Mori}, {Perri}, {Pivovaroff}, {Puccetti}, {Rana}, {Westergaard}, {Willis},
  {Zoglauer}, {An}, {Bachetti}, {Barri{\`e}re}, {Bellm}, {Bhalerao},
  {Brejnholt}, {Fuerst}, {Liebe}, {Markwardt}, {Nynka}, {Vogel}, {Walton},
  {Wik}, {Alexander}, {Cominsky}, {Hornschemeier}, {Hornstrup}, {Kaspi},
  {Madejski}, {Matt}, {Molendi}, {Smith}, {Tomsick}, {Ajello}, {Ballantyne},
  {Balokovi{\'c}}, {Barret}, {Bauer}, {Blandford}, {Brandt}, {Brenneman},
  {Chiang}, {Chakrabarty}, {Chenevez}, {Comastri}, {Dufour}, {Elvis}, {Fabian},
  {Farrah}, {Fryer}, {Gotthelf}, {Grindlay}, {Helfand}, {Krivonos}, {Meier},
  {Miller}, {Natalucci}, {Ogle}, {Ofek}, {Ptak}, {Reynolds}, {Rigby},
  {Tagliaferri}, {Thorsett}, {Treister}, \& {Urry}}]{2013ApJ...770..103H}
{Harrison}, F.~A., {Craig}, W.~W., {Christensen}, F.~E., {et~al.} 2013, \apj,
  770, 103, \dodoi{10.1088/0004-637X/770/2/103}

\bibitem[{{Heckman} {et~al.}(2005){Heckman}, {Ptak}, {Hornschemeier}, \&
  {Kauffmann}}]{2005ApJ...634..161H}
{Heckman}, T.~M., {Ptak}, A., {Hornschemeier}, A., \& {Kauffmann}, G. 2005,
  \apj, 634, 161, \dodoi{10.1086/491665}

\bibitem[{{Hern{\'a}n-Caballero} {et~al.}(2015){Hern{\'a}n-Caballero},
  {Alonso-Herrero}, {Hatziminaoglou}, {Spoon}, {Ramos Almeida}, {D{\'\i}az
  Santos}, {H{\"o}nig}, {Gonz{\'a}lez-Mart{\'\i}n}, \&
  {Esquej}}]{2015ApJ...803..109H}
{Hern{\'a}n-Caballero}, A., {Alonso-Herrero}, A., {Hatziminaoglou}, E.,
  {et~al.} 2015, \apj, 803, 109, \dodoi{10.1088/0004-637X/803/2/109}

\bibitem[{{Hickox} \& {Alexander}(2018)}]{2018ARA&A..56..625H}
{Hickox}, R.~C., \& {Alexander}, D.~M. 2018, \araa, 56, 625,
  \dodoi{10.1146/annurev-astro-081817-051803}

\bibitem[{{Hopkins} {et~al.}(2006){Hopkins}, {Hernquist}, {Cox}, {Di Matteo},
  {Robertson}, \& {Springel}}]{2006ApJS..163....1H}
{Hopkins}, P.~F., {Hernquist}, L., {Cox}, T.~J., {et~al.} 2006, \apjs, 163, 1,
  \dodoi{10.1086/499298}

\bibitem[{{Hunter}(2007)}]{2007CSE.....9...90H}
{Hunter}, J.~D. 2007, Computing in Science and Engineering, 9, 90,
  \dodoi{10.1109/MCSE.2007.55}

\bibitem[{{Jansen} {et~al.}(2001){Jansen}, {Lumb}, {Altieri}, {Clavel}, {Ehle},
  {Erd}, {Gabriel}, {Guainazzi}, {Gondoin}, {Much}, {Munoz}, {Santos},
  {Schartel}, {Texier}, \& {Vacanti}}]{2001A&A...365L...1J}
{Jansen}, F., {Lumb}, D., {Altieri}, B., {et~al.} 2001, \aap, 365, L1,
  \dodoi{10.1051/0004-6361:20000036}

\bibitem[{{Kammoun} {et~al.}(2019){Kammoun}, {Miller}, {Zoghbi}, {Oh}, {Koss},
  {Mushotzky}, {Brenneman}, {Brandt}, {Proga}, {Lohfink}, {Kaastra}, {Barret},
  {Behar}, \& {Stern}}]{2019ApJ...877..102K}
{Kammoun}, E.~S., {Miller}, J.~M., {Zoghbi}, A., {et~al.} 2019, \apj, 877, 102,
  \dodoi{10.3847/1538-4357/ab1c5f}

\bibitem[{{Khachikian} \& {Weedman}(1974)}]{1974ApJ...192..581K}
{Khachikian}, E.~Y., \& {Weedman}, D.~W. 1974, \apj, 192, 581,
  \dodoi{10.1086/153093}

\bibitem[{{Kocevski} {et~al.}(2015){Kocevski}, {Brightman}, {Nandra},
  {Koekemoer}, {Salvato}, {Aird}, {Bell}, {Hsu}, {Kartaltepe}, {Koo}, {Lotz},
  {McIntosh}, {Mozena}, {Rosario}, \& {Trump}}]{2015ApJ...814..104K}
{Kocevski}, D.~D., {Brightman}, M., {Nandra}, K., {et~al.} 2015, \apj, 814,
  104, \dodoi{10.1088/0004-637X/814/2/104}

\bibitem[{{Kormendy} \& {Ho}(2013)}]{2013ARA&A..51..511K}
{Kormendy}, J., \& {Ho}, L.~C. 2013, \araa, 51, 511,
  \dodoi{10.1146/annurev-astro-082708-101811}

\bibitem[{{LaMassa} {et~al.}(2012){LaMassa}, {Heckman}, \&
  {Ptak}}]{2012ApJ...758...82L}
{LaMassa}, S.~M., {Heckman}, T.~M., \& {Ptak}, A. 2012, \apj, 758, 82,
  \dodoi{10.1088/0004-637X/758/2/82}

\bibitem[{{LaMassa} {et~al.}(2019){LaMassa}, {Yaqoob}, {Boorman}, {Tzanavaris},
  {Levenson}, {Gandhi}, {Ptak}, \& {Heckman}}]{2019ApJ...887..173L}
{LaMassa}, S.~M., {Yaqoob}, T., {Boorman}, P.~G., {et~al.} 2019, \apj, 887,
  173, \dodoi{10.3847/1538-4357/ab552c}

\bibitem[{{Lanzuisi} {et~al.}(2015){Lanzuisi}, {Perna}, {Delvecchio}, {Berta},
  {Brusa}, {Cappelluti}, {Comastri}, {Gilli}, {Gruppioni}, {Mignoli}, {Pozzi},
  {Vietri}, {Vignali}, \& {Zamorani}}]{2015A&A...578A.120L}
{Lanzuisi}, G., {Perna}, M., {Delvecchio}, I., {et~al.} 2015, \aap, 578, A120,
  \dodoi{10.1051/0004-6361/201526036}

\bibitem[{{Lebouteiller} {et~al.}(2011){Lebouteiller}, {Barry}, {Spoon},
  {Bernard-Salas}, {Sloan}, {Houck}, \& {Weedman}}]{2011ApJS..196....8L}
{Lebouteiller}, V., {Barry}, D.~J., {Spoon}, H.~W.~W., {et~al.} 2011, \apjs,
  196, 8, \dodoi{10.1088/0067-0049/196/1/8}

\bibitem[{{Maiolino} {et~al.}(1998){Maiolino}, {Salvati}, {Bassani}, {Dadina},
  {della Ceca}, {Matt}, {Risaliti}, \& {Zamorani}}]{1998A&A...338..781M}
{Maiolino}, R., {Salvati}, M., {Bassani}, L., {et~al.} 1998, \aap, 338, 781.
\newblock \doarXiv{astro-ph/9806055}

\bibitem[{{Marconi} {et~al.}(2004){Marconi}, {Risaliti}, {Gilli}, {Hunt},
  {Maiolino}, \& {Salvati}}]{2004MNRAS.351..169M}
{Marconi}, A., {Risaliti}, G., {Gilli}, R., {et~al.} 2004, \mnras, 351, 169,
  \dodoi{10.1111/j.1365-2966.2004.07765.x}

\bibitem[{{Marshall} {et~al.}(1980){Marshall}, {Boldt}, {Holt}, {Miller},
  {Mushotzky}, {Rose}, {Rothschild}, \& {Serlemitsos}}]{1980ApJ...235....4M}
{Marshall}, F.~E., {Boldt}, E.~A., {Holt}, S.~S., {et~al.} 1980, \apj, 235, 4,
  \dodoi{10.1086/157601}

\bibitem[{{Moshir} \& {et al.}(1990)}]{1990IRASF.C......0M}
{Moshir}, M., \& {et al.} 1990, IRAS Faint Source Catalogue, 0

\bibitem[{{Murphy} \& {Yaqoob}(2009)}]{2009MNRAS.397.1549M}
{Murphy}, K.~D., \& {Yaqoob}, T. 2009, \mnras, 397, 1549,
  \dodoi{10.1111/j.1365-2966.2009.15025.x}

\bibitem[{{Nandra} {et~al.}(2007){Nandra}, {O'Neill}, {George}, \&
  {Reeves}}]{2007MNRAS.382..194N}
{Nandra}, K., {O'Neill}, P.~M., {George}, I.~M., \& {Reeves}, J.~N. 2007,
  \mnras, 382, 194, \dodoi{10.1111/j.1365-2966.2007.12331.x}

\bibitem[{{Netzer}(2015)}]{2015ARA&A..53..365N}
{Netzer}, H. 2015, \araa, 53, 365, \dodoi{10.1146/annurev-astro-082214-122302}

\bibitem[{{Planck Collaboration} {et~al.}(2020){Planck Collaboration},
  {Aghanim}, {Akrami}, {Ashdown}, {Aumont}, {Baccigalupi}, {Ballardini},
  {Banday}, {Barreiro}, {Bartolo}, {Basak}, {Battye}, {Benabed}, {Bernard},
  {Bersanelli}, {Bielewicz}, {Bock}, {Bond}, {Borrill}, {Bouchet}, {Boulanger},
  {Bucher}, {Burigana}, {Butler}, {Calabrese}, {Cardoso}, {Carron},
  {Challinor}, {Chiang}, {Chluba}, {Colombo}, {Combet}, {Contreras}, {Crill},
  {Cuttaia}, {de Bernardis}, {de Zotti}, {Delabrouille}, {Delouis}, {Di
  Valentino}, {Diego}, {Dor{\'e}}, {Douspis}, {Ducout}, {Dupac}, {Dusini},
  {Efstathiou}, {Elsner}, {En{\ss}lin}, {Eriksen}, {Fantaye}, {Farhang},
  {Fergusson}, {Fernandez-Cobos}, {Finelli}, {Forastieri}, {Frailis},
  {Fraisse}, {Franceschi}, {Frolov}, {Galeotta}, {Galli}, {Ganga},
  {G{\'e}nova-Santos}, {Gerbino}, {Ghosh}, {Gonz{\'a}lez-Nuevo}, {G{\'o}rski},
  {Gratton}, {Gruppuso}, {Gudmundsson}, {Hamann}, {Handley}, {Hansen},
  {Herranz}, {Hildebrandt}, {Hivon}, {Huang}, {Jaffe}, {Jones}, {Karakci},
  {Keih{\"a}nen}, {Keskitalo}, {Kiiveri}, {Kim}, {Kisner}, {Knox},
  {Krachmalnicoff}, {Kunz}, {Kurki-Suonio}, {Lagache}, {Lamarre}, {Lasenby},
  {Lattanzi}, {Lawrence}, {Le Jeune}, {Lemos}, {Lesgourgues}, {Levrier},
  {Lewis}, {Liguori}, {Lilje}, {Lilley}, {Lindholm}, {L{\'o}pez-Caniego},
  {Lubin}, {Ma}, {Mac{\'\i}as-P{\'e}rez}, {Maggio}, {Maino}, {Mandolesi},
  {Mangilli}, {Marcos-Caballero}, {Maris}, {Martin}, {Martinelli},
  {Mart{\'\i}nez-Gonz{\'a}lez}, {Matarrese}, {Mauri}, {McEwen}, {Meinhold},
  {Melchiorri}, {Mennella}, {Migliaccio}, {Millea}, {Mitra},
  {Miville-Desch{\^e}nes}, {Molinari}, {Montier}, {Morgante}, {Moss}, {Natoli},
  {N{\o}rgaard-Nielsen}, {Pagano}, {Paoletti}, {Partridge}, {Patanchon},
  {Peiris}, {Perrotta}, {Pettorino}, {Piacentini}, {Polastri}, {Polenta},
  {Puget}, {Rachen}, {Reinecke}, {Remazeilles}, {Renzi}, {Rocha}, {Rosset},
  {Roudier}, {Rubi{\~n}o-Mart{\'\i}n}, {Ruiz-Granados}, {Salvati}, {Sandri},
  {Savelainen}, {Scott}, {Shellard}, {Sirignano}, {Sirri}, {Spencer},
  {Sunyaev}, {Suur-Uski}, {Tauber}, {Tavagnacco}, {Tenti}, {Toffolatti},
  {Tomasi}, {Trombetti}, {Valenziano}, {Valiviita}, {Van Tent}, {Vibert},
  {Vielva}, {Villa}, {Vittorio}, {Wandelt}, {Wehus}, {White}, {White},
  {Zacchei}, \& {Zonca}}]{2020A&A...641A...6P}
{Planck Collaboration}, {Aghanim}, N., {Akrami}, Y., {et~al.} 2020, \aap, 641,
  A6, \dodoi{10.1051/0004-6361/201833910}

\bibitem[{{Richstone} \& {Schmidt}(1980)}]{1980ApJ...235..361R}
{Richstone}, D.~O., \& {Schmidt}, M. 1980, \apj, 235, 361,
  \dodoi{10.1086/157640}

\bibitem[{{Risaliti} {et~al.}(1999){Risaliti}, {Maiolino}, \&
  {Salvati}}]{1999ApJ...522..157R}
{Risaliti}, G., {Maiolino}, R., \& {Salvati}, M. 1999, \apj, 522, 157,
  \dodoi{10.1086/307623}

\bibitem[{{Roche} \& {Aitken}(1985)}]{1985MNRAS.215..425R}
{Roche}, P.~F., \& {Aitken}, D.~K. 1985, \mnras, 215, 425,
  \dodoi{10.1093/mnras/215.3.425}

\bibitem[{{Salim} {et~al.}(2007){Salim}, {Rich}, {Charlot}, {Brinchmann},
  {Johnson}, {Schiminovich}, {Seibert}, {Mallery}, {Heckman}, {Forster},
  {Friedman}, {Martin}, {Morrissey}, {Neff}, {Small}, {Wyder}, {Bianchi},
  {Donas}, {Lee}, {Madore}, {Milliard}, {Szalay}, {Welsh}, \&
  {Yi}}]{2007ApJS..173..267S}
{Salim}, S., {Rich}, R.~M., {Charlot}, S., {et~al.} 2007, \apjs, 173, 267,
  \dodoi{10.1086/519218}

\bibitem[{{Skrutskie} {et~al.}(2006){Skrutskie}, {Cutri}, {Stiening},
  {Weinberg}, {Schneider}, {Carpenter}, {Beichman}, {Capps}, {Chester},
  {Elias}, {Huchra}, {Liebert}, {Lonsdale}, {Monet}, {Price}, {Seitzer},
  {Jarrett}, {Kirkpatrick}, {Gizis}, {Howard}, {Evans}, {Fowler}, {Fullmer},
  {Hurt}, {Light}, {Kopan}, {Marsh}, {McCallon}, {Tam}, {Van Dyk}, \&
  {Wheelock}}]{2006AJ....131.1163S}
{Skrutskie}, M.~F., {Cutri}, R.~M., {Stiening}, R., {et~al.} 2006, \aj, 131,
  1163, \dodoi{10.1086/498708}

\bibitem[{{Tanimoto} {et~al.}(2019){Tanimoto}, {Ueda}, {Odaka}, {Kawaguchi},
  {Fukazawa}, \& {Kawamuro}}]{2019ApJ...877...95T}
{Tanimoto}, A., {Ueda}, Y., {Odaka}, H., {et~al.} 2019, \apj, 877, 95,
  \dodoi{10.3847/1538-4357/ab1b20}

\bibitem[{{Tanimoto} {et~al.}(2022){Tanimoto}, {Ueda}, {Odaka}, {Yamada}, \&
  {Ricci}}]{2022ApJS..260...30T}
{Tanimoto}, A., {Ueda}, Y., {Odaka}, H., {Yamada}, S., \& {Ricci}, C. 2022,
  \apjs, 260, 30, \dodoi{10.3847/1538-4365/ac5f59}

\bibitem[{{Terashima} {et~al.}(2015){Terashima}, {Hirata}, {Awaki}, {Oyabu},
  {Gandhi}, {Toba}, \& {Matsuhara}}]{2015ApJ...814...11T}
{Terashima}, Y., {Hirata}, Y., {Awaki}, H., {et~al.} 2015, \apj, 814, 11,
  \dodoi{10.1088/0004-637X/814/1/11}

\bibitem[{{Toba} {et~al.}(2020){Toba}, {Yamada}, {Ueda}, {Ricci}, {Terashima},
  {Nagao}, {Wang}, {Tanimoto}, \& {Kawamuro}}]{2020ApJ...888....8T}
{Toba}, Y., {Yamada}, S., {Ueda}, Y., {et~al.} 2020, \apj, 888, 8,
  \dodoi{10.3847/1538-4357/ab5718}

\bibitem[{{Turner} {et~al.}(1997){Turner}, {George}, {Nandra}, \&
  {Mushotzky}}]{1997ApJ...488..164T}
{Turner}, T.~J., {George}, I.~M., {Nandra}, K., \& {Mushotzky}, R.~F. 1997,
  \apj, 488, 164, \dodoi{10.1086/304701}

\bibitem[{{Ueda} {et~al.}(2014){Ueda}, {Akiyama}, {Hasinger}, {Miyaji}, \&
  {Watson}}]{2014IAUS..304..125U}
{Ueda}, Y., {Akiyama}, M., {Hasinger}, G., {Miyaji}, T., \& {Watson}, M.~G.
  2014, in Multiwavelength AGN Surveys and Studies, ed. A.~M. {Mickaelian} \&
  D.~B. {Sanders}, Vol. 304, 125--131, \dodoi{10.1017/S1743921314003536}

\bibitem[{{Urry} \& {Padovani}(1995)}]{1995PASP..107..803U}
{Urry}, C.~M., \& {Padovani}, P. 1995, \pasp, 107, 803, \dodoi{10.1086/133630}

\bibitem[{{Woo} \& {Urry}(2002)}]{2002ApJ...579..530W}
{Woo}, J.-H., \& {Urry}, C.~M. 2002, \apj, 579, 530, \dodoi{10.1086/342878}

\bibitem[{{Wright} {et~al.}(2010){Wright}, {Eisenhardt}, {Mainzer}, {Ressler},
  {Cutri}, {Jarrett}, {Kirkpatrick}, {Padgett}, {McMillan}, {Skrutskie},
  {Stanford}, {Cohen}, {Walker}, {Mather}, {Leisawitz}, {Gautier}, {McLean},
  {Benford}, {Lonsdale}, {Blain}, {Mendez}, {Irace}, {Duval}, {Liu}, {Royer},
  {Heinrichsen}, {Howard}, {Shannon}, {Kendall}, {Walsh}, {Larsen}, {Cardon},
  {Schick}, {Schwalm}, {Abid}, {Fabinsky}, {Naes}, \&
  {Tsai}}]{2010AJ....140.1868W}
{Wright}, E.~L., {Eisenhardt}, P. R.~M., {Mainzer}, A.~K., {et~al.} 2010, \aj,
  140, 1868, \dodoi{10.1088/0004-6256/140/6/1868}

\bibitem[{{Xu} {et~al.}(2020){Xu}, {Sun}, {Xue}, {Li}, \&
  {He}}]{2020RAA....20..147X}
{Xu}, J., {Sun}, M.-Y., {Xue}, Y.-Q., {Li}, J.-Y., \& {He}, Z.-C. 2020,
  Research in Astronomy and Astrophysics, 20, 147,
  \dodoi{10.1088/1674-4527/20/9/147}

\end{thebibliography}
\bibliographystyle{aasjournal}



\end{document}